\documentclass[fleqn,usenatbib]{mnras}

\usepackage{newtxtext,newtxmath}
\usepackage[T1]{fontenc}

\DeclareRobustCommand{\VAN}[3]{#2}
\let\VANthebibliography\thebibliography
\def\thebibliography{\DeclareRobustCommand{\VAN}[3]{##3}\VANthebibliography}

\usepackage{graphicx}	% Including figure files
\usepackage{amsmath}	% Advanced maths commands

\title[Silane--Methane Competition in Sub-Neptunes]{Silane--Methane Competition  in Sub-Neptune Atmospheres as a Diagnostic of Metallicity and Magma Oceans}

\author[K. Hakim et al.]{
Kaustubh Hakim$^{1,2}$\thanks{E-mail: kaustubh.hakim@kuleuven.be},
Dan J. Bower$^{3}$,
Fabian L. Seidler$^{3}$
and Paolo A. Sossi$^{3}$
\\
$^{1}$ KU Leuven, Institute of Astronomy, Celestijnenlaan 200D, 3001 Leuven, Belgium \\
$^{2}$ Royal Observatory of Belgium, Ringlaan 3, 1180 Brussels, Belgium  \\
$^{3}$ ETH Zurich, Department of Earth and Planetary Sciences, Clausiusstrasse 25, 8092 Zurich, Switzerland
}

\date{Accepted XXX. Received YYY; in original form ZZZ}

\pubyear{2025}

\begin{document}
\label{firstpage}
\pagerange{\pageref{firstpage}--\pageref{lastpage}}
\maketitle

% Abstract of the paper
\begin{abstract}

The James Webb Space Telescope is characterising the atmospheres of sub-Neptunes. The presence of magma oceans on sub-Neptunes is expected to strongly alter the chemistry of their envelopes and observable atmospheres. At the magma ocean–envelope boundary (MEB, $>$10~kbar), gas properties deviate from ideality, yet the effects of real gas behaviour on chemical equilibria remain underexplored. Here, we compute equilibrium between magma--gas and gas--gas reactions using real gas equations of state in the H--He--C--N--O--Si system for TOI-421b, a canonical hot sub-Neptune potentially hosting a magma ocean. We find that H and N are the most soluble in magma, followed by He and C. We fit real gas equations of state to experimental data on SiH$_4$, and show that, for a fully molten mantle, SiH$_4$ dominates at the MEB under accreted gas metallicity of 1$\times$ solar, but is supplanted by CH$_4$ at 100$\times$ solar. Lower mantle melt fractions lower both magma-derived Si abundances in the envelope and the solubility of H and He in magma, yielding H$_2$- and He-rich envelopes. Projecting equilibrium chemistry through the observable atmosphere (1~mbar--100~bar), we find that `clouds' of Si-bearing condensates strongly deplete Si-bearing gases, although SiH$_4$ remains key, especially when a solar gas is accreted. SiH$_4$/CH$_4$ and Si/C ratios increase with mantle melt fraction and decrease with accreted gas metallicity. The competition between SiH$_4$ and CH$_4$ is therefore diagnostic of metallicity and presence of magma oceans on sub-Neptunes with equilibrium temperatures below 1000~K. The corollary is that H$_2$- and He-rich, SiH$_4$-deficient and CH$_4$-bearing observable atmospheres may indicate a limited role or absence of magma oceans on sub-Neptunes.

\end{abstract}

% Select between one and six entries from the list of approved keywords.
% Don't make up new ones.

%%%%%%%%%%%%%%%%%%%%%%%%%%%%%%%%%%%%%%%%%%%%%%%%%%

%%%%%%%%%%%%%%%%% BODY OF PAPER %%%%%%%%%%%%%%%%%%

\section{Introduction} \label{sec:intro}

The size demographics of transiting exoplanets indicate that the sub-Neptune population peaks around $\sim$2.5~$R_{\oplus}$ and exhibits a sharp decrease from $2.5R_{\oplus}$  to $3.5 R_{\oplus}$ \citep{2017AJ....154..109F,2018MNRAS.479.4786V,2020AJ....160..108B}. Mass-radius determinations of transiting exoplanets reveal that sub-Neptunes ($1.8 R_{\oplus} < R < 3.5 R_{\oplus}$) possess a range of bulk densities consistent with a combination of silicates and/or water and an H mass fraction on the order of 0.1--1\% of planet mass (0.1--1 wt\%~H) and/or condensed H$_2$O up to $\sim$10~wt\% \citep{2014ApJ...792....1L,2019PNAS..116.9723Z,2022Sci...377.1211L}. Although the James Webb Space Telescope's (JWST) transmission spectroscopy measurements of the observable atmospheres of sub-Neptunes reveal one or two carbon-bearing molecules (CO, CO$_2$ or CH$_4$) with no H$_2$O \citep[K2-18b,][]{2023ApJ...956L..13M}, potential H$_2$O \citep[TOI-270d,][]{2024arXiv240303325B}, or some H$_2$O \citep[TOI-421b,][]{2025ApJ...984L..44D}, the composition of their envelopes (bulk gas/fluid layers) is still unconstrained. Yet, determining their chemical composition and volatile mass fraction is key to understanding planet formation and migration pathways of small exoplanets \citep{2021JGRE..12606639B,2024NatAs...8..463B}.

Sub-Neptunes are expected to lose their accretionary heat over Gyr timescales because of their greenhouse envelopes with a thickness on the order of the Earth's radius \citep{2013ApJ...775...80F,2018ApJ...869..163V,2025ApJ...988L..55W}. \cite{2025ApJ...989...28T} demonstrate that, for hot sub-Neptunes with incident stellar radiation more than 100$\times$ of that received by Earth ($>100 F_{\oplus}$), or equivalently a planet equilibrium temperature $T_{\rm eq}>900$~K, and an H mass fraction of $\sim$1~wt\%, their silicate mantle likely remains fully molten (global magma ocean) with a magma--envelope boundary (MEB) temperature $\sim$3000~K at an evolutionary age of $\sim$10~Gyr. Magma--envelope interactions are expected to produce detectable signatures \citep{2020ApJ...891..111K,2022PSJ.....3..127S,2023A&A...671A.138Z,2023MNRAS.524..981M,2024A&A...691A.159S}, making hot sub-Neptunes ideal candidates for interior--atmosphere characterisation \citep[e.g.,][]{2022MNRAS.513.4015G}.

What sets the chemical abundances of elements in the envelopes of sub-Neptunes? Astronomical perspectives suggest that sub-Neptunes capture substantial fractions of H$_2$/He-rich nebular gas of the order 1~\% of their total mass \citep{2013ApJ...777...34M, 2022MNRAS.513.4015G}. Thus, the envelope of sub-Neptunes likely begins with stellar metallicity. The evolutionary effects of atmospheric escape processes such as photoevaporation and core-powered mass loss suggest the loss of a significant fraction of the H$_2$/He envelope with age \citep{2018MNRAS.476..759G,2024MNRAS.528.1615O}. Because lighter species, namely H$_2$ and He, are lost preferentially to heavier ones (e.g., CO, H$_2$O), the abundances of elements heavier than H and He increase with age, resulting in a higher metallicity than the stellar value. The empirical planet mass--metallicity trends, stemming from solar system planets, imply a metallicity of around 100$\times$ solar for planets around the mass of Neptune \citep{2017Sci...356..628W,2019ApJ...887L..20W}. Photoevaporation can also drive a differential loss of elements heavier than H and He, further complicating the determination of metallicity \citep{2025ApJ...991..121L}. Given uncertainties including the measurement of exoplanet host star metallicity, there is little consensus on the metallicity of sub-Neptunes from recent observations, with interpretations spanning the range between 1--100$\times$ solar \citep{2023ApJ...956L..13M,2024arXiv240303325B,2025ApJ...984L..44D,2025Sci...390S3660L}.

Recent geochemical modelling efforts include the effects of a magma ocean on the composition of the envelope  \citep[e.g.,][]{2020ApJ...891..111K,2022PSJ.....3...93B, 2022PSJ.....3..127S,2023A&A...674A.224C, 2023MNRAS.524..981M,2025ApJ...988L..55W,2025ApJ...983...97C,2025ApJ...987..174I,2025arXiv250700499B}. In this regime, magma--gas reactions dictate the elemental abundances in the envelope. Magma-gas reactions produce Si-bearing gases, especially SiH$_4$ and SiO \citep{2023MNRAS.524..981M,2023A&A...674A.224C}. Because the atmospheric spectra of
Si-bearing gases are detectable \citep{2023MNRAS.524..981M,2023A&A...671A.138Z,2025ApJ...987..174I}, it is crucial to consider all key chemical interactions at the magma--envelope boundary. The solubility of hydrogen (H$_2$ and H$_2$O) in magma allows the storage of large amounts of hydrogen in the interior of sub-Neptunes \citep{2018ApJ...854...21C,2019ApJ...887L..33K}. 
The dissolution of hydrogen in magma can potentially explain the slope of the radius cliff, a sharp decrease in the population of sub-Neptunes from 2.5--3.5~$R_{\rm \oplus}$ in the exoplanet size distribution \citep[][]{2019ApJ...887L..33K}, although its effect is likely more modest than previously thought \citep{2022PSJ.....3..127S}. 
The solubility of H$_2$O further modifies the atmospheric composition of Si-bearing gases \citep{STB23, 2025ApJ...987..174I}. Since H$_2$ and H$_2$O are also coupled through gas-phase reactions at equilibrium, H$_2$ and H$_2$O cannot dissolve independently of each other. The real gas behaviour of volatile gases and gas solubility can result in hybrid atmospheres, a mixture of outgassed and primordial atmospheres \citep{2024ApJ...963..157T,2025ApJ...994...28H}. However, none of the existing models \citep{2022PSJ.....3..127S,2023A&A...674A.224C, 2023MNRAS.524..981M, 2024ApJ...975..101R, 2024ApJ...975...14S,2025ApJ...987..174I} consider the real gas behaviour of Si-bearing gases in governing their partial pressures in the envelopes of sub-Neptunes.

In this work, we model the envelope composition at the magma--envelope boundary ($>$10~kbar) and observable atmospheric composition (1~mbar--100~bar) of a canonical hot sub-Neptune, TOI-421b, by varying H mass fraction, metallicity, and mantle melt fraction in the H--He--O--C--N--Si chemical system using \texttt{Atmodeller} \citep{2025arXiv250700499B}. We simultaneously model chemical reactions between gas and magma species, with comprehensive geochemical datasets on gas solubility and real gas behaviour of all considered gases, including SiO and SiH$_4$. We evaluate the effects of magma--envelope coupling on the equilibrium atmospheric composition under SiO$_2$ condensation. We demonstrate the impact of the metallicity of accreted gas and the mantle melt fraction in controlling the SiH$_4$/CH$_4$ and Si/C ratios in the observable atmospheres of sub-Neptunes. 

\begin{figure}
	\includegraphics[width=\columnwidth]{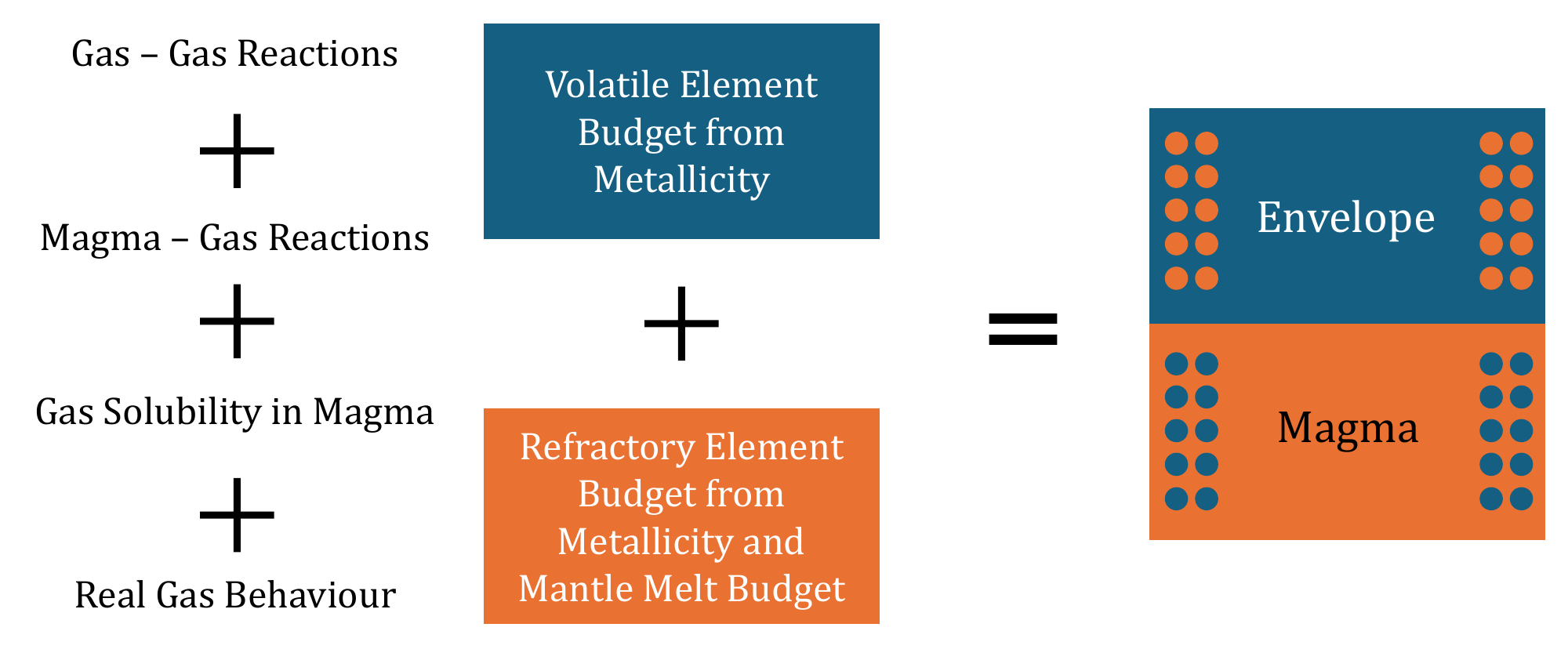}
    \caption{Illustration of magma--envelope chemical coupling. Volatile element budgets are set by the metallicity of accreted gas, and the refractory element budgets are set by the sum of the metallicity of accreted gas and their budget in the bulk silicate Earth scaled by planet mass. Gas--gas reactions, magma--gas reactions, gas solubilities in magma and real gas behaviour are modelled using \texttt{Atmodeller} \citep{2025arXiv250700499B}.  }
    \label{fig:illustration}
\end{figure}

\section{Model} \label{sec:methods}

\subsection{Model Setup and Elemental Mass Constraints}\label{sec:methodsOverview}

The methods presented here are implemented in \texttt{Atmodeller}, an open-source Python package that models the atmospheric chemistry of rocky planets and sub-Neptunes with self-consistent interior--atmosphere coupling \citep{2025arXiv250700499B}. We apply our methods to the hot sub-Neptune, TOI-421b, with radius $R_P = 2.64 R_{\oplus}$, mass $M_P = 6.7 M_{\oplus}$ and equilibrium temperature $T_{\rm eq}=920$~K \citep{2024A&A...686A.301K}. Considering the envelope mass to be negligible compared to the planet mass, the radius of MEB $R_{\rm MEB}$ is calculated from the Earth-scaled mass--radius relation, $R_{\rm MEB} = 1.02 M_{P}^{0.252} = 1.65 R_{\oplus}$ \citep{2018Icar..313...61H}. The MEB temperature of a highly irradiated (100$F_{\oplus}$), $\sim5 M_{\oplus}$ planet after 10~Gyr of evolution should be $T_{\rm MEB}\sim3000$~K, high enough to keep the mantle almost fully molten, i.e., mantle melt fraction of unity \citep{2025ApJ...989...28T}. MEB temperatures can be sensitive to the envelope mass fraction, although MEB temperatures between 2500--3500~K have a limited impact on the results and do not affect our conclusions. Henceforth, we fix $T_{\rm MEB}=3000$~K. We also consider mantle melt fractions of lower than 100\%, down to 1\%. Because the extent of metallic core–-silicate mantle equilibration in sub-Neptunes is uncertain, we instead consider a range of accreted gas metallicities to approximate the C/H and N/H fractionation expected during core formation \citep{2024GeCoA.376..100H}. We note that more oxidising conditions are possible than modelled in this work \citep{2019Sci...365..903A,2022GeCoA.328..221H}.

\begin{table}
\centering
\caption{Elemental mass constraints for the models in the H--He--C--N--O--Si system at 3000~K} \label{tab:massconstraints}
\begin{tabular}{r|ll|ll}
\hline
\hline
Metallicity & \multicolumn2c{1$\times$ solar} & \multicolumn2c{100$\times$ solar} \\
H mass fraction & 0.1~wt\%~H & 1~wt\%~H & 0.1~wt\%~H & 1~wt\%~H  \\
\hline
\multicolumn4c{Mass constraints for mantle melt fraction of 100\%} \\
He [wt\%] & 0.033 & 0.33  & 0.033 & 0.33 \\
C [wt\%] & 0.00029 & 0.0029 & 0.029 & 0.29 \\
N [wt\%] & 0.000095 & 0.00095 & 0.0095 & 0.095 \\
Si [wt\%] & 14.5 & 14.5 & 14.51 & 14.6 \\
O [wt\%] & 16.5 & 16.51 & 16.59 & 17.35 \\
\hline
\multicolumn4c{Mass constraints for mantle melt fraction of 10\%} \\
He [wt\%] & 0.033 & 0.33  & 0.033 & 0.33 \\
C [wt\%] & 0.00029 & 0.0029 & 0.029 & 0.29 \\
N [wt\%] & 0.000095 & 0.00095 & 0.0095 & 0.095 \\
Si [wt\%] & 1.45 & 1.451 & 1.46 & 1.54 \\
O [wt\%] & 1.651 & 1.659 & 1.735 & 2.5 \\
\hline
\multicolumn4c{Mass constraints for mantle melt fraction of 1\%} \\
He [wt\%] & 0.033 & 0.33 & 0.033 & 0.33 \\
C [wt\%] & 0.00029 & 0.0029 & 0.029 & 0.29 \\
N [wt\%] & 0.000095 & 0.00095 & 0.0095 & 0.095 \\
Si [wt\%] & 0.1451 & 0.146 & 0.154 & 0.24 \\
O [wt\%] & 0.1659 & 0.1735 & 0.25 & 1.02 \\
\hline
\end{tabular}
\smallskip \\
{Note: All wt\% values imply elemental mass per cent of planet mass. The solar metallicity values are from \citet{2009LanB...4B..712L} and the SiO$_2$ value for bulk silicate Earth is from \citet{2014Palme}.  }
\end{table}

We model magma--envelope coupling by considering gas-gas, magma-gas and gas solubility reactions, and real gas behaviour (Fig.~\ref{fig:illustration}). Although chemical reactions at the magma--envelope boundary result in the exchange of elements between the magma and the envelope, the total elemental budget is conserved.
The budget of accreted H is a free parameter (0.1--1\% of planet mass, i.e., 0.1--1~wt\%). The budgets of volatile elements other than H (here He, C and N) are defined by the product of the metallicity of accreted gas and the mass fraction of accreted H (Table~\ref{tab:massconstraints}). For example, for solar metallicity and 1~wt\%~H, the budgets of He, C, and N are 0.33~wt\%, 0.0029~wt\%, and 0.00095~wt\%, respectively, where the solar metallicity values are obtained from \citet[][their Table~8]{2009LanB...4B..712L}. For 100$\times$ solar metallicity and 1~wt\%~H, the He budget is the same as for solar metallicity, but the budgets of C and N are scaled up by a factor of 100 (Table~\ref{tab:massconstraints}). The budgets of the `refractory' elements, Si and O, are the sum of (a) the product of the metallicity constraint and the H mass fraction, and (b) their budget in the mantle as liquid SiO$_2$. The mantle mass of SiO$_2$ is scaled from its value in the bulk silicate Earth to that of TOI-421b. Specifically, for a mantle melt fraction of unity, the mantle budget of Si is 14.5\% of planet mass, i.e., 14.5~wt\% and the budget of O is 16.5~wt\%, derived from the bulk silicate Earth's SiO$_2$ of 45.4~wt\% \citep{2014Palme}. SiO$_2$ is expected to be the primary component of the magma and the focus of the study. The remainder of the mantle is treated as chemically inert. By adding the available Si and O budgets for solar metallicity to their mantle budgets, the total Si and O budgets remain similar to their mantle budgets. However, for 100$\times$ solar metallicity and 1~wt\%~H, the total Si and O budgets increase slightly to 14.6~wt\% and 17.35~wt\%, respectively (Table~\ref{tab:massconstraints}). When the mantle melt fraction is reduced to 10\% and 1\%, the mantle Si and O budgets are scaled down by a factor of 10 and 100, respectively (Table~\ref{tab:massconstraints}).

\subsection{Real Gas Behaviour} \label{sec:methodsRealGas}

The total MEB pressures of sub-Neptunes exceed 10~kbar, above which the pressure--volume--temperature $P-V-T$ properties of gases strongly deviate from the ideal gas assumption. To capture the correct thermodynamic behaviour of real gases, the fugacity coefficient $\phi$, usually derived from a relevant equation of state (EOS), provides the real gas correction factor to convert the partial pressure $P_{i}$ of gas species $i$ to fugacity $f_{i}$ ($\phi_i=1$ in the ideal gas regime)
\begin{align} 
f_i = \phi_i P_i.
\end{align}
We model the real gas behaviour of all gases based on standard equations of state (Sects.~\ref{sec:methodsHOSi} and \ref{sec:methodsHHeCNOSi}). For SiO and SiH$_4$, we describe below our approach to model their real gas behaviour. 

% \vspace{5mm}
% \noindent \textit{Real gas behaviour of SiO}
% \vspace{2mm}

\subsubsection{SiO}

\noindent For SiO$_{\rm (g)}$, following \cite{2016JGRE..121.1641C}, we employ the law of corresponding states, where the properties of a given gas species can be described with respect to their reduced temperatures ($T_r$) and pressures ($P_r$);
\begin{align} 
    T_r = \frac{T}{T_c}, \\
    P_r = \frac{P}{P_c},
\end{align}
where $T_c=3431$~K and $P_c=4544$~bar are the critical temperature and pressure of SiO, respectively \citep{2016JGRE..121.1641C}. This states that, at the same reduced temperature and reduced pressure, the compressibility factor of any gas is the same, where
\begin{equation}
Z_c = \frac{P_c V_c}{R T_c}.
\end{equation}
The modelled value of $Z_c$ depends on the chosen form of the EOS, with the simplest, van der Waals EOS yielding $Z_c$ = 0.375, and the Redlich-Kwong EOS giving $Z_c$ = 0.333. Because experimentally determined values of $Z_c$ range from $\sim$0.23 to 0.31, we combine the law of corresponding states with the Redlich-Kwong EOS \citep{redlichkwong1949,2016JGRE..121.1641C}:
\begin{equation}
   P = \frac{RT}{V - b} - \frac{a}{\sqrt{T} V (V + b)} 
\end{equation}
in which the solutions for $a$ and $b$ can be found by simultaneously solving the first and second derivatives of pressure with respect to volume at the critical point, $\left(\frac{\partial P}{\partial V}\right)_{T=T_c}$ and $\left(\frac{\partial^2 P}{\partial V^2}\right)_{T=T_c}$,
\begin{align}
V_c &= \frac{RT_c}{\mathbf{3} P_c}, \\
a &= \frac{RT_c^{3/2} V_c}{3(2^{1/3} - 1)}, \\
b &= (2^{1/3} - 1) V_c,
\end{align}
with $a = 4.48433 \times 10^{-5}$~m$^6$~bar~K$^{0.5}$~mol$^{-2}$ and b = $5.43922 \times 10^{-6}$~m$^3$~mol$^{-1}$. 

% \vspace{5mm}
% \noindent \textit{Real gas behaviour of SiH$_4$} 
% \vspace{2mm}

\subsubsection{SiH$_4$}

\noindent There are no published high-pressure high-temperature equations of state for silane SiH$_4$. Existing models for the atmospheric compositions of sub-Neptunes highlight the stability of silane in reducing, H-rich atmospheres \citep{2023A&A...674A.224C,2023MNRAS.524..981M,2025ApJ...987..174I}, yet none consider its non-ideality in the gas phase. Earlier efforts to define a $P-V-T$ EOS for SiH$_4$, in the absence of experimental data at the time, assumed compressibilities similar to those of CH$_4$ by applying the law of corresponding states \citep{2011PhRvB..83n4102S}. However, accurate $P$ and $T$ ranges for the extrapolation of compressibilities based on critical properties are $P$ $\sim$ 10$P_c$ and $T$ $\sim$ 2--3 $T_c$ \citep{reid1987properties}, or roughly 500~bar and 800~K, respectively, for silane \citep[$T_c$ = 269.7~K; $P_c$ = 48.4~bar,][]{reid1987properties}, which are insufficient for our purposes. 

Here, we fit recent $P-V-T$ data for fluid silane obtained between 44.2$\pm$0.4~GPa and 1200$\pm$30~K up to 138.7$\pm$0.7~GPa and 4109$\pm$80~K via shock compression \citep{2018PhyB..541...89W} to a second-order (quadratic) Virial EOS, in which the compressibility factor $Z$ is expressed in terms of $P$ and $T$,
\begin{equation}
    Z(P,T) = 1 + A(T) P + B(T) P^2,
    \label{eq:Virial_SiH4}
\end{equation}
where $A(T)$ and $B(T)$ are temperature-dependent Virial coefficients parameterised as:
\begin{align}
    A(T) &= A_0 + A_1T + A_2T^2, \\
    B(T) &= B_0 + B_1T + B_2T^2,
    \label{eq:Virial_SiH4_AB}
\end{align}
and the molar volume ($V$) is related to $Z$ via
\begin{equation}
    V(P,T) = \frac{Z(P,T)RT}{P}.
    \label{eq:Z_volume}
\end{equation}

We perform an error-weighted non-linear least squares fit by minimising the objective function:
\begin{equation}
\chi^2 = \sum_{i=1}^{N} \left( \frac{V_{\text{model}}(P_i, T_i; \boldsymbol{\theta}) - V_{\text{obs},i}}{\sigma_{V_{\text{obs},i}}} \right)^2
\end{equation}
across all 14 data points, $i$, and the associated error in their volumes \citep[$\sigma_{V_{{\rm obs},i}}$,][their Table~4]{2018PhyB..541...89W}. Because the uncertainties in $P$, $V$ and $T$ are strongly correlated in shock experiments, we assume perfect covariance among them and account only for the errors in volume. Equivalent results could be obtained by considering uncertainties in either $P$ or $T$. Uncertainties in the fitted parameters, $A_0$, $A_1$, $A_2$, $B_0$, $B_1$, $B_2$, collectively represented by $\boldsymbol{\theta}$, were determined from the covariance matrix;
\begin{equation}
\mathbf{C} = \chi^2_r \cdot (\mathbf{J}^\mathrm{T} \mathbf{J})^{-1}
\end{equation}
where $\mathbf{J}$ is the Jacobian matrix of partial derivatives and $\chi^2_r$ is the reduced chi-squared estimate of the residual variance. The 1$\sigma$ uncertainties for each coefficient are given by the square root of the corresponding diagonal entry of \textbf{C}. A fit to a cubic Virial EOS ($Z = 1 + A(T)P + B(T)P^2 + C(T)P^3$) in which the temperature-dependent terms take the form $A(T) = A_0 + A_1T$ was also attempted but was found to yield a slightly poorer fit to the data.

\begin{figure}
    \centering
    \includegraphics[width=1\linewidth]{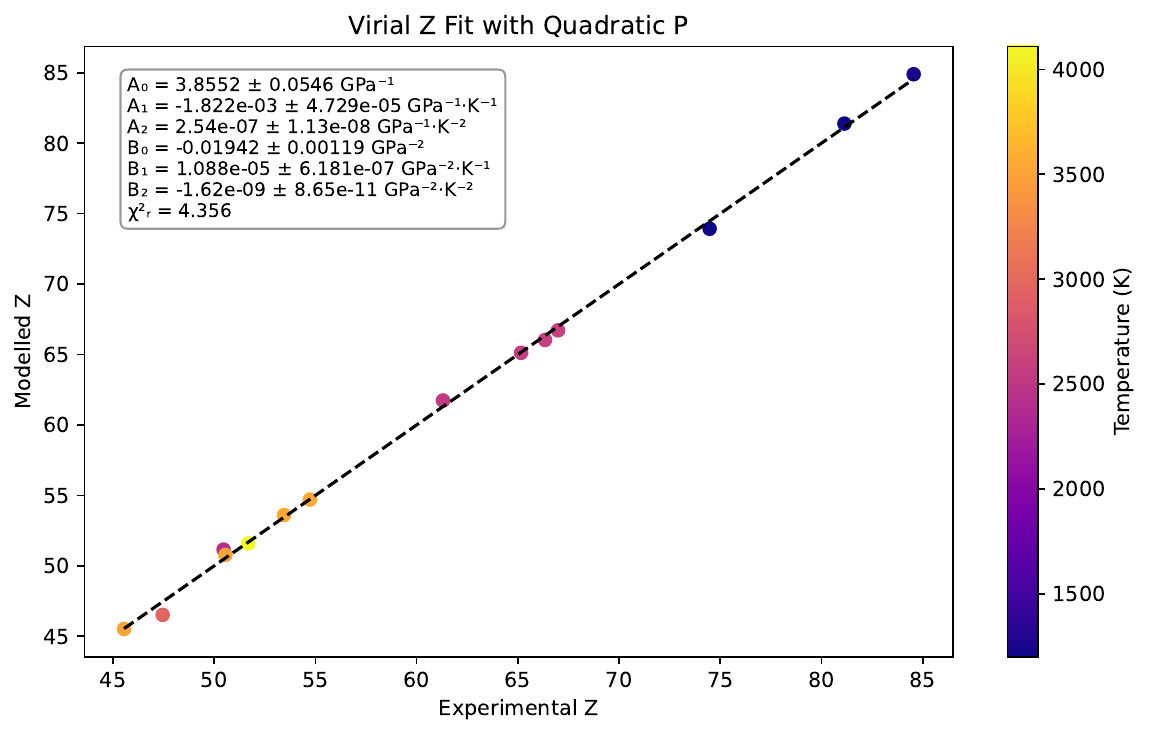}
    \caption{Comparison of the compressibility factors ($Z$) modelled in Eq. \ref{eq:Virial_SiH4} with those observed in the shock experiments of \citet{2018PhyB..541...89W}. The data are colour-coded by temperature (in K). The derived coefficients, their associated uncertainties, and the reduced chi-squared ($\chi^2_r$) values are shown in the legend.}
    \label{fig:SiH4_fit}
\end{figure}

The best-fit coefficients for SiH$_4$ and their uncertainties are reported in Table \ref{tab:virial_params} and Fig. \ref{fig:SiH4_fit}. The resulting fugacity coefficient at 3000~K and 5~GPa is 28.4$\pm$2.1, similar to that of CH$_4$ at identical conditions, as might be expected from their similar $T_c$ and $P_c$. The fluid silane data are calibrated for $P > 44$~GPa; however, the fugacity coefficient extrapolates correctly to unity at $ P = 1$~bar.

\begin{table}
\centering
\caption{Fitted parameters of the quadratic Virial equation of state (eqs. \ref{eq:Virial_SiH4}-\ref{eq:Virial_SiH4_AB}) for SiH$_4$ and their 1$\sigma$ uncertainties.}
\begin{tabular}{lll}
\hline
Parameter & Value & Units \\
\hline
$A_0$ & $3.8552 \pm 0.0546$ & GPa$^{-1}$ \\
$A_1$ & $-1.822 \times 10^{-3} \pm 4.729 \times 10^{-5}$ & GPa$^{-1}$·K$^{-1}$ \\
$A_2$ & $2.54 \times 10^{-7} \pm 1.13 \times 10^{-8}$ & GPa$^{-1}$·K$^{-2}$ \\
$B_0$ & $-0.01942 \pm 0.00119$ & GPa$^{-2}$ \\
$B_1$ & $1.088 \times 10^{-5} \pm 6.181 \times 10^{-7}$ & GPa$^{-2}$·K$^{-1}$ \\
$B_2$ & $-1.62 \times 10^{-9} \pm 8.65 \times 10^{-11}$ & GPa$^{-2}$·K$^{-2}$ \\
\hline
\end{tabular}
\label{tab:virial_params}
\end{table}

Although we refer to the equation of state of SiH$_4$ in our model,  at higher pressures and temperatures, speciation changes of the type:
\begin{equation}
    \mathrm{SiH_{n(g)} \rightleftharpoons \frac{(n-1)}{2}H_{2(g)} + SiH_{n-1(g)}}
    \label{eq:dehydrogenation}
\end{equation}
where $n \leq 4$, or formation of oligosilanes
\begin{equation}
    \mathrm{(SiH_n)_{y(g)} \rightleftharpoons ySiH_{n(g)}}
    \label{eq:oligosilane}
\end{equation}
may occur. Insofar as only the stoichiometry of SiH$_4$ is fixed in the shock experiments of \cite{2018PhyB..541...89W}, our EOS recovers the behaviour of ``SiH$_4$'', even if its speciation likely diverges from that of silane at extreme $P-T$.

\subsection{The H--O--Si Chemical System} \label{sec:methodsHOSi}

First, we demonstrate our methods using the H--O--Si chemical system. Then, we extend it to the H--He--C--N--O--Si chemical system. The H--O--Si system is a minimalist chemical representation of the magma--envelope coupling because H is expected to be the dominant element in the envelope, Si is a key element in the silicate magma, and O is present both in the magma and the envelope and plays a crucial role in redox reactions.

\begin{table*}
\centering
\caption{Real gas equations of state (EOS) of key gases in the H--He--C--N--O--Si chemical system} \label{tab:realEOS}
\begin{tabular}{lllrcc}
\hline
\hline
\multicolumn3c{} & \multicolumn2c{Calibrated Range}\\
Species & EOS & Reference & Pressure (GPa) & Temperature (K)  \\
\hline
H$_2$ & Interpolated ab initio data & \cite{2021ApJ...917....4C}   & 0--10$^{13}$ & 100--$10^8$ \\
He & Interpolated ab initio data & \cite{2021ApJ...917....4C} & 0--10$^{13}$ & 100--$10^8$ \\
H$_2$O & Compensated Redlich-Kwong & \cite{HP91} &  0--5 & 400--1700 \\ 
O$_2$ & Virial Cubic & \cite{SS92} & 0--100 & 200-3000 \\
CH$_4$ & Virial Cubic & \cite{SS92} & 0--100 & 200--3000 \\
CO & Virial Cubic & \cite{SS92} & 0--100 & 200--3000 \\
CO$_2$ & Virial Cubic & \cite{SS92} & 0--100 & 200--3000 \\
N$_2$ & Virial Cubic & \cite{SF87} & 0--100 & 200--3000 \\
NH$_3$ &  Redlich-Kwong & \cite{reid1987properties} &  0--0.01 & 300--500 \\
SiO &  Redlich-Kwong & \cite{2016JGRE..121.1641C} & 0--5 & 1000--10000 \\
SiH$_4$ & Virial Quadratic & \cite{2018PhyB..541...89W}, This work & 0--138.7 & 1200-4100 \\
\hline
\end{tabular}
\end{table*}

\begin{table*}
\centering
\caption{Chemical reactions in the H--He--C--N--O--Si chemical system} \label{tab:reactions}
\begin{tabular}{rclcl}
\hline
\hline
\multicolumn3c{Chemical Reactions} & Reaction Constant & Reference  \\
\hline
\multicolumn{3}{l}{Gas-Gas Reactions} \\
\(\rm{H_{2(g)} + 0.5\,O_{2(g)}}\) & \(\rightleftharpoons\) & \(\rm{H_2O_{(g)}}\) & $K_1 = \frac{f_{\rm{H_2O_{(g)}}}}{f_{\rm{H_{2(g)}}} \sqrt{f_{\rm{O_{2(g)}}}}}$ & \citet{MZG02} \\
\(\rm{CO_{(g)} + 0.5\,O_{2(g)}}\) & \(\rightleftharpoons\) & \(\rm{CO_{2(g)}}\) & $K_2 = \frac{f_{\rm{CO_{2(g)}}}}{f_{\rm{CO_{(g)}}} \sqrt{f_{\rm{O_{2(g)}}}}}$ & \citet{MZG02}  \\
\(\rm{CO_{(g)} + 2\,H_{2(g)}}\) & \(\rightleftharpoons\) & \(\rm{CH_{4(g)} + 0.5\,O_{2(g)}}\) & $K_3 = \frac{f_{\rm{CH_{4(g)}}} \sqrt{f_{\rm{O_{2(g)}}}}}{f_{\rm{CO_{(g)}}} f^2_{\rm{H_{2(g)}}} }$ & \citet{MZG02} \\
\(\rm{N_{2(g)} + 3\,H_{2(g)}}\) & \(\rightleftharpoons\) & \(\rm{2\,NH_{3(g)}}\) & $K_4 = \frac{f_{\rm{N_{2(g)}}}}{f^3_{\rm{H_{2(g)}}} f^2_{\rm{NH_{3(g)}}}}$ & \citet{MZG02} \\
\(\rm{SiH_{4(g)} + 0.5\,O_{2(g)}}\) & \(\rightleftharpoons\) & \(\rm{SiO_{(g)} + 2\,H_{2(g)}}\) & $K_5 = \frac{f_{\rm{SiO_{(g)}}} f^2_{\rm{H_{2(g)}}} }{f_{\rm{SiH_{4(g)}}}  \sqrt{f_{\rm{O_{2(g)}}}}}$ & \citet{MZG02} \\
\\
\multicolumn{3}{l}{Magma-Gas Reaction} \\
\(\rm{SiO_{2(l)}}\) & \(\rightleftharpoons\) & \(\rm{SiO_{(g)}  + 0.5\,O_{2(g)}}\) & $K_6 = \frac{f_{\rm{SiO_{(g)}}} \sqrt{f_{\rm{O_{2(g)}}}} }{a_{\rm{SiO_{2(l)}}}}$ & \citet{MZG02} \\
\\
\multicolumn{3}{l}{Gas Dissolution Reactions in Magma} \\
\(\rm{H_{2(g)}}\) & \(\rightleftharpoons\) & \(\rm{H_{2(m)}}\) & $K_7 = \frac{a_{\rm{H_{2(m)}}}}{f_{\rm{H_{2(g)}}}}$ & \citet[their Table~2]{HWA12} \\
\(\rm{H_{2}O_{(g)} + O^{2-}_{(m)}}\) & \(\rightleftharpoons\) & \(\rm{2\,OH^-_{(m)}}\) & $K_8 = \frac{a^2_{\rm{OH^-_{(m)}}}}{f_{\rm{H_2O_{(g)}}} a_{\rm{O^{2-}_{(m)}}}}$  &  \citet{STB23} \\
\(\rm{He_{(g)}}\) & \(\rightleftharpoons\) & \(\rm{He_{(m)}}\) & $K_9 = \frac{a_{\rm{He_{(m)}}}}{f_{\rm{He_{(g)}}}}$  & \citet{JWB86} \\
\(\rm{CH_{4(g)}}\) & \(\rightleftharpoons\) & \(\rm{CH_{4(m)}}\) & $K_{10} = \frac{a_{\rm{CH_{4(m)}}}}{f_{\rm{CH_{4(g)}}}}$  & \citet[their Eq. 7a and 8]{AHW13} \\
\(\rm{CO_{(g)}}\) & \(\rightleftharpoons\) & \(\rm{CO_{(m)}}\) & $K_{11} = \frac{a_{\rm{CO_{(m)}}}}{f_{\rm{CO_{(g)}}}}$  & \citet{YNN19} \\
\(\rm{CO_{2(g)} + O^{2-}_{(m)}}\) & \(\rightleftharpoons\) & \(\rm{CO^{2-}_{3(m)}}\) & $K_{12} = \frac{a_{\rm{CO^{2-}_{3(m)}}}}{f_{\rm{CO_{2(g)}}} a_{\rm{O^{2-}_{(m)}}}}$  & \citet[their Eq.~6]{DSH95} \\
\(\rm{N_{2(g)}}\) & \(\rightleftharpoons\) & \(\rm{N_{2(m)}}\) & $K_{13} = \frac{a_{\rm{N_{2(m)}}}}{f_{\rm{N_{2(g)}}}}$  & \citet[their Eq.~23]{LMH03} \\
\hline
\end{tabular}
\end{table*}

\subsubsection{Chemical Reactions with Real Gases} \label{sec:methodsReactions}

In the H--O--Si chemical system, the following set of gas--gas reactions is chosen, which involve gases H$_2$, H$_2$O, O$_2$, SiH$_4$ and SiO.
\begin{align} 
\rm{H_{2(g)} + 0.5 O_{2(g)}} &\rightleftharpoons \rm{H_2O_{(g)}} \label{eq:H2H2O} \\
\rm{SiH_{4(g)} + 0.5 O_{2(g)}} &\rightleftharpoons \rm{SiO_{(g)} + 2 H_{2(g)}} 
\end{align}
We model the real gas behaviour (Table~\ref{tab:realEOS}) of H$_2$ \citep[ab initio equations compilation,][]{2021ApJ...917....4C}, H$_2$O \citep[compensated Redlich-Kwong EOS,][]{1998JMetG..16..309H}, O$_2$ \citep[cubic virial EOS,][]{SS92}, SiH$_4$ \citep[quadratic virial EOS,][This work]{2018PhyB..541...89W} and SiO \citep[Redlich-Kwong EOS,][]{2016JGRE..121.1641C}. The fugacities of the five gases listed above constitute the five unknowns. The two reactions result in two equations, where the equilibrium constant of each reaction is equal to the product of the fugacities of the reaction products divided by the product of the fugacities of the reactants (Table~\ref{tab:reactions}). The equilibrium constants are obtained from \citet{MZG02}. Combining two equations with equilibrium constants and the budgets of O and Si (Sect.~\ref{sec:methodsOverview}), the set of equations can be exactly solved for the given H mass fraction by adhering to mass balance.

Liquid SiO$_2$ is the primary component of silicate magma. In addition to the previous two gas--gas reactions, the following magma--gas reaction becomes important. 
\begin{align} 
\rm{SiO_{2(l)}} &\rightleftharpoons \rm{SiO_{(g)} + 0.5 O_{2(g)}}.
\label{eq:SiO2}
\end{align}
This additional equation introduces another unknown, the stability of SiO$_{\rm 2(l)}$. \texttt{Atmodeller} evaluates the stability of SiO$_{\rm 2(l)}$ and calculates the mass fractions of Si in magma and envelope by applying the extended law of mass action \citep{2025arXiv250700499B}.  In all magma--envelope models discussed in this paper, we find that SiO$_{\rm 2(l)}$ is stable (i.e., activity of SiO$_{\rm 2(l)}$ $a_{\rm SiO_{2(l)}}$ = 1), which therefore maintains consistency with our assumption of a magma ocean. The equilibrium constant for this reaction is obtained from \citet{MZG02}. 

Magma oceans can dissolve gases to an extent that modifies the atmospheric composition. We utilise the latest data for H$_2$O solubility in primitive magma \citep[peridotite melt,][]{STB23} and H$_2$ solubility in an evolved magma  \citep[basaltic melt,][]{2012E&PSL.345...38H} as listed in Table~\ref{tab:reactions}. In the H--O--Si chemical system, the two reactions governing gas solubilities in magma can be written as: 
\begin{align} 
\rm{H_{2(g)}} &\rightleftharpoons \rm{H_{2(m)}} \\
\rm{H_2O_{(g)}} + \rm{O^{2-}_{(m)}} &\rightleftharpoons \rm{2OH^-_{(m)}}.
\end{align}

We note here that, in our model, the dissolution of these volatile components in the magma does not affect the activity of SiO$_2$, which is evaluated to be unity. Insofar as these volatile-bearing components (predominantly OH$^-$) comprise only up to a few per cent (by weight) of the mantle mass, neglecting their effect on $a_{\rm SiO_{2(l)}}$ should introduce minimal uncertainty on mass and composition of the computed envelopes.

\begin{figure}
	\includegraphics[width=\columnwidth]{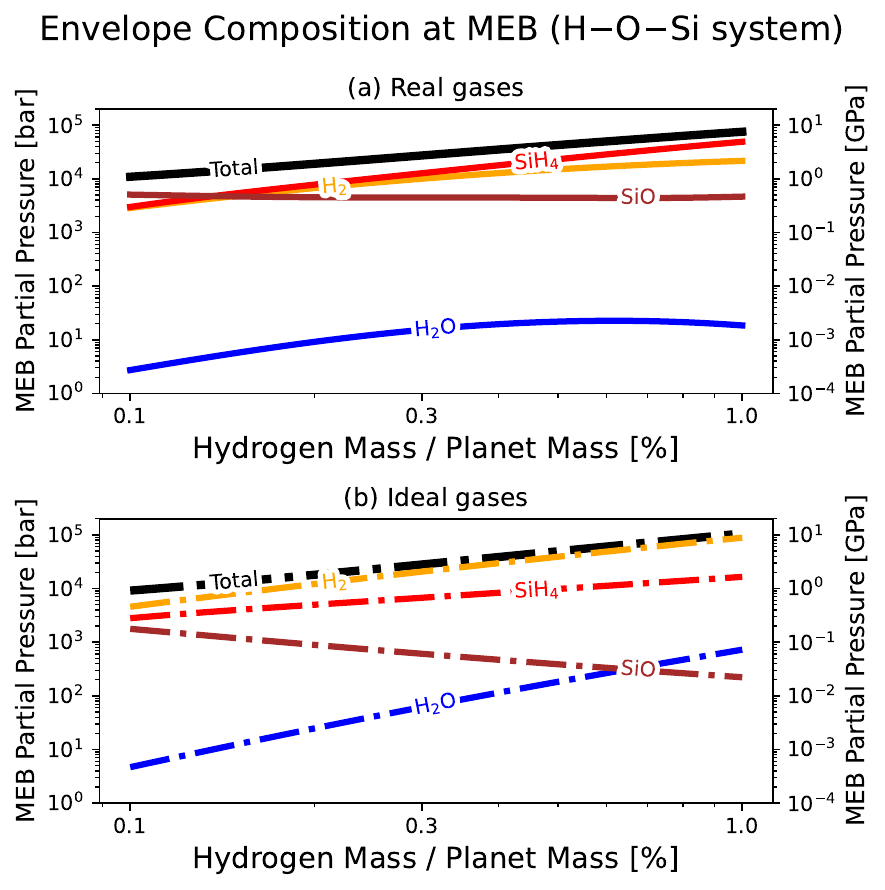}
    \caption{Partial pressures of H$_2$, H$_2$O, SiO and SiH$_4$ at the magma--envelope boundary (MEB) of TOI-421b ($R_{\rm MEB} = 1.65 R_{\oplus}$, $R_P = 2.64 R_{\oplus}$, $M_P = 6.7 M_{\oplus}$, $T_{\rm MEB}=3000$~K with a fully molten mantle) in the H--O--Si chemical system. (a) Envelope composition with real gas behaviour and gas solubility. (b) Envelope composition with ideal gas behaviour and gas solubility.  O$_2$ partial pressures are less than 10$^{-7}$~bar and not shown.  }
    \label{fig:HOSi_model}
\end{figure}

\begin{figure}
	\includegraphics[width=0.95\columnwidth]{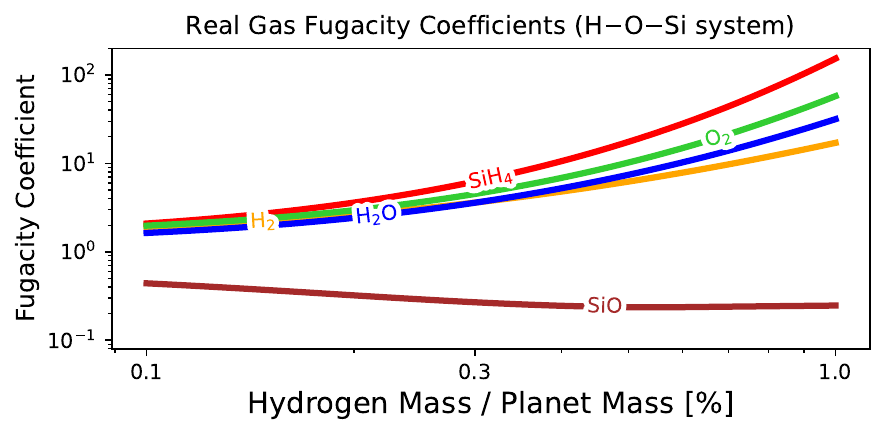}
    \caption{Fugacity coefficients of the H--O--Si gases for the real gas case of Fig.~\ref{fig:HOSi_model}. }
    \label{fig:HOSi_model_fugacity}
\end{figure}

\begin{figure}
\includegraphics[width=0.95\columnwidth]{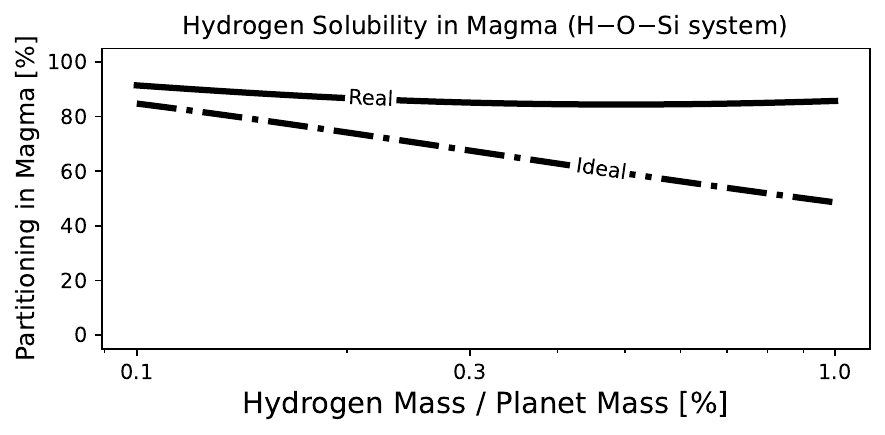}
    \caption{The solubility of H (in the form of H$_2$ and H$_2$O) as a function of the total H budget relative to the planet mass for real and ideal gas cases from Fig.~\ref{fig:HOSi_model}. }
    \label{fig:HOSi_model_solubility}
\end{figure}

\subsubsection{Envelope Composition with Real Gases and Solubility} \label{sec:methodsHOSiResults}

In the H--O--Si chemical system with real gases, solubility of H$_2$ and H$_2$O in magma and elemental budgets of 14.5~wt\% Si and 16.5~wt\% O, for 0.1--1~wt\%~H, the MEB total pressure ranges between 11--75~kbar (1.1--7.5~GPa, Fig.~\ref{fig:HOSi_model}(a)). We find that the MEB envelope composition of TOI-421b is dominated by SiO at H mass fractions below 0.15~wt\% and by SiH$_4$ for $> 0.15$~wt\%~H. Partial pressures of H$_2$O are nearly 3--4 orders of magnitude lower than those of H$_2$. The predominance of Si-bearing gases in the envelope results from the vaporisation of SiO$_{\rm 2(l)}$ (Sect.~\ref{sec:methodsHOSi}).

Ignoring real gas behaviour under the ideal gas assumption, the total MEB pressure ranges between 9--106~kbar (0.9--10.6~GPa) for 0.1--1~wt\%~H (Fig.~\ref{fig:HOSi_model}(b)). Compared to the real gas case, H$_2$, H$_2$O and SiH$_4$ partial pressures are higher and the SiO partial pressure is lower. This is because the real gas fugacity coefficients of SiO are lower than unity and those of SiH$_4$, H$_2$ and H$_2$O are higher than unity (Fig.~\ref{fig:HOSi_model_fugacity}). Excluding real gas behaviour results in a twofold effect. First, the solubilities of both H$_2$ and H$_2$O in magma are lower by up to a factor of one-third when the real gas behaviour is excluded (Fig.~\ref{fig:HOSi_model_solubility}). For the ideal case, this is clearly visible in the higher total MEB pressure at 1 wt\% H, which is due to the lower partitioning of H (Fig.~\ref{fig:HOSi_model}(b)). This is because the solubility of a gas is proportional to its fugacity, not its partial pressure. In the real case, higher fugacities of H$_2$ and H$_2$O lead to a higher partitioning of H in magma, thereby decreasing the total MEB pressure (Fig.~\ref{fig:HOSi_model}(a)).

Second, the contrasting nature of the fugacity coefficients of Si-bearing gases, $\phi_{\rm SiO} < 1$ and $\phi_{\rm SiH_4} > 1$  results in their MEB partial pressures diverging more from each other under the ideal gas assumption (Fig.~\ref{fig:HOSi_model}(b)) compared to the real gas case (Fig.~\ref{fig:HOSi_model}(a)). This also results in a lower total MEB pressure at 0.1~wt\%~H in the ideal case than in the real case.  Because $T_c$ of SiO$_{\rm (g)}$ is much higher than that of SiH$_4$ (and other gases considered), $\phi_{\rm SiO}$ decreases (albeit mildly), indicating that the critical compressibility $Z_c$ is $<$1. This behaviour occurs because $T/T_c$ for SiO is much lower than for the other gases, and hence the `$a$' term (the attractive term), which in the Redlich-Kwong EOS depends on $T_c^{3/2}$, dominates over the `$b$' term (the repulsive term) over a wider pressure range. Consequently, the volume of SiO$_{\rm (g)}$ is lower than that expected from the ideal gas law under the same pressure and temperature. In contrast, the compressibility of SiH$_4$ leads to the opposing effect, where volumes are higher than those of an ideal gas. The result is that $P_{\rm SiO}$/$P_{\rm SiH_4}$ can exceed unity at low wt\%~H in the real case, but remains below unity down to 0.1 ~wt\%~H in the ideal case.

\begin{figure*}
	\includegraphics[width=\textwidth]{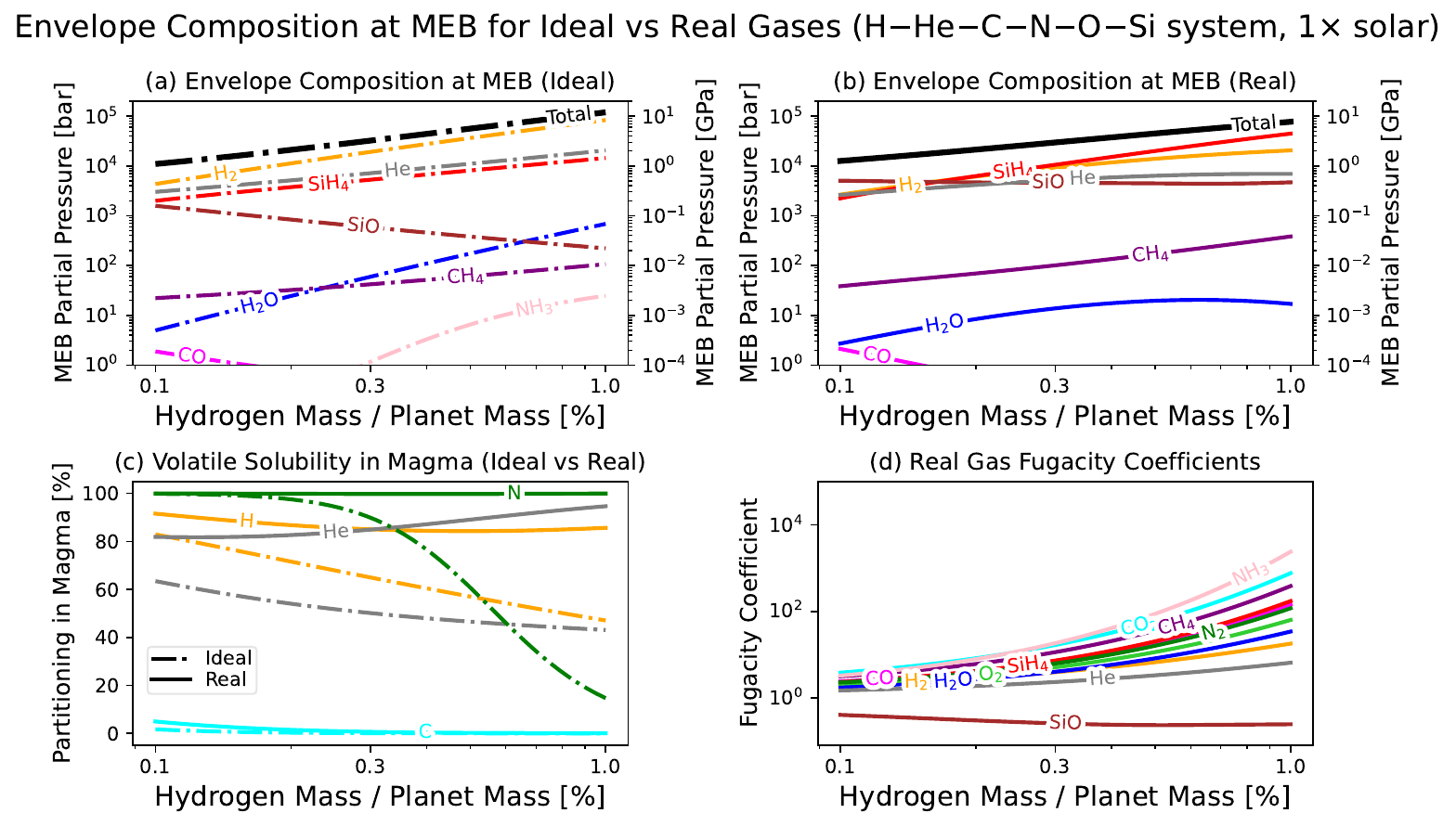}
    \caption{Envelope composition at the MEB of TOI-421b ($R_{\rm MEB} = 1.65 R_{\oplus}$, $R_P = 2.64 R_{\oplus}$, $M_P = 6.7 M_{\oplus}$, $T_{\rm MEB}=3000$~K with a fully molten mantle) as a function of total H budget relative to planet mass in the H--He--C--N--O--Si chemical system. Elemental budgets are defined by solar metallicity (Table~\ref{tab:massconstraints}). Reactions, gas solubility laws, and real gas equations are described in Sect.~\ref{sec:methods}. (a) MEB partial pressures of the considered gases assuming ideal gas behaviour. O$_2$ partial pressures range between 10$^{-7}$--10$^{-9}$ bar (fugacities between $-$4.3 and $-$6 log$_{10}$ units below the iron-wüstite (IW) buffer). (b) MEB partial pressures of the considered gases under real gas behaviour. O$_2$ partial pressures range between 8$\times$10$^{-10}$ and 7$\times$10$^{-11}$ bar (oxygen fugacities between IW$-$6.2 and IW$-$5.7). Full range of MEB partial pressures is shown in Fig.~\ref{fig:full_model_ideal_real_fullrange}. (c) Element partitioning in magma of H, He, C and N for ideal (a) and real (b) cases. (d) Real gas fugacity coefficients of all gases in (b). }
    \label{fig:full_model_ideal_real}
\end{figure*}

\subsection{The H--He--C--N--O--Si Chemical System} \label{sec:methodsHHeCNOSi} 

We extend our methods to the H--He--C--N--O--Si chemical system by choosing the gases, H$_2$, H$_2$O, He, CH$_4$, CO, CO$_2$, NH$_3$, N$_2$, SiH$_4$ and SiO, and liquid SiO$_2$ representing magma. All chemical reactions are listed in Table~\ref{tab:reactions}. The real gas equations of state are listed in Table~\ref{tab:realEOS}. In our formulation, H mass fraction, mantle melt fraction and metallicity of accreted gas are the three free parameters that control the budgets of volatile and refractory elements (Table~\ref{tab:massconstraints}). We calculate the envelope composition at the MEB of TOI-421b ($T_{\rm MEB}=3000$~K) for H mass fractions between 0.1--1~\% of planet mass and mantle melt fractions between 1--100\%. The elemental budgets are described in Sect.~\ref{sec:methodsOverview}.

\begin{figure*}
	\includegraphics[width=\textwidth]{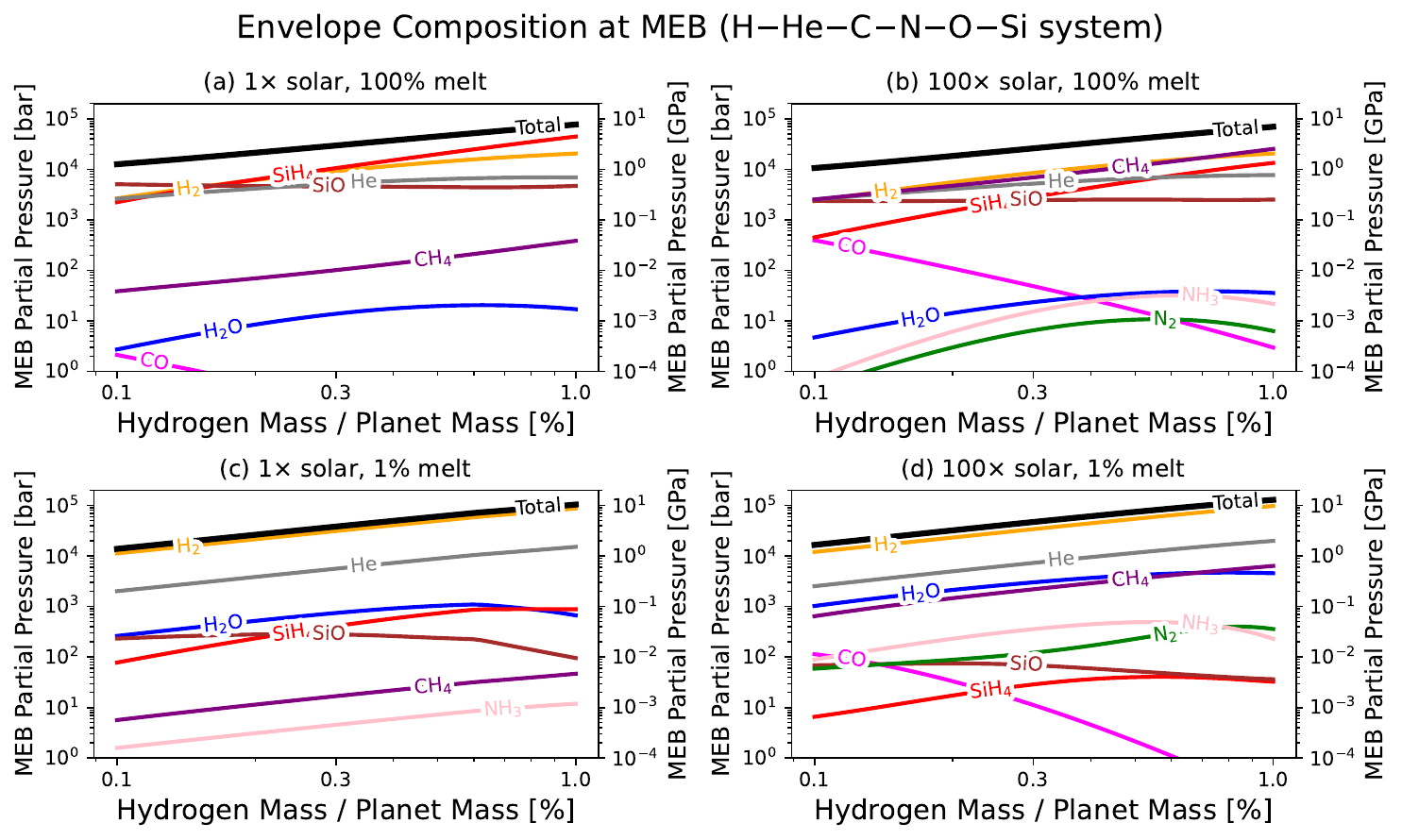}
    \caption{Envelope composition in the H--He--C--N--O--Si chemical system of TOI-421b ($R_{\rm MEB} = 1.65 R_{\oplus}$, $R_P = 2.64 R_{\oplus}$, $M_P = 6.7 M_{\oplus}$, $T_{\rm MEB}=3000$~K) as a function of H mass fraction. Elemental budgets, reactions, gas solubility laws, and real gas equations are described in Sect.~\ref{sec:methods}. MEB partial pressures of the considered gases for (a) 1$\times$ solar metallicity and 100\% mantle melt (IW--6.2 to IW--5.7), (b) 100$\times$ solar metallicity and 100\% mantle melt (IW--5.5 to IW--5.2), (c) 1$\times$ solar metallicity and 1\% mantle melt (IW--3.4 to IW--3.1), and (d) 100$\times$ solar metallicity and 1\% mantle melt (IW--2.3 to IW--1.6). Full range of MEB partial pressures, including that of O$_2$, is shown in Fig.~\ref{fig:full_model_envelope_fullrange}. The corresponding partitioning of volatile elements in magma and real gas fugacity coefficients are given in Fig.~\ref{fig:full_model_solubility} and Fig.~\ref{fig:full_model_fugcoeff}, respectively. }
    \label{fig:full_model_envelope}
\end{figure*}

\section{Envelopes and Atmospheres with Magma--Envelope Coupling} \label{sec:results}

\subsection{Effects of Solubility and Real Gases} \label{sec:resultsEnvelope1}

We compare the MEB envelope compositions for the ideal and real gas cases of TOI-421b in the H--He--C--N--O--Si chemical system with mantle melt fraction of 100\% and accreted gas metallicity of 1$\times$ solar (Sect.~\ref{sec:methodsHHeCNOSi}). For 0.1--1~wt\% H, MEB total pressures range between 13--77~kbar (1.3--7.7~GPa) for the real gas case and between 11--120~kbar (1.1--12~GPa) for the ideal gas case (Fig.~\ref{fig:full_model_ideal_real}(a,b)). In contrast to the ideal case with H$_2$ being the dominant gas (Fig.~\ref{fig:full_model_ideal_real}(a)), the real gas case results in SiO or SiH$_4$ being the dominant gases (Fig.~\ref{fig:full_model_ideal_real}(b)).

There are two main reasons for the differences between the real and ideal gas cases. First, the fugacity coefficient of SiO is below unity, which is in contrast to other considered gases (Fig.~\ref{fig:full_model_ideal_real}(d)). This effect results in $P_{\rm SiO} > P_{\rm SiH_4}$ in the real gas case between 0.1--0.15~wt\%~H, in contrast to the ideal gas case (Fig.~\ref{fig:full_model_ideal_real}(a,b)). Because of this, a larger amount of SiO is vaporised in the real case, increasing the Si abundance in the envelope, especially at low H mass fractions.

Second, a larger fraction of H partitions into magma in the real gas case due to the higher solubility of H$_2$ and H$_2$O compared to the ideal gas case (Fig.~\ref{fig:full_model_ideal_real}(c)). This effect is most pronounced at the highest considered H mass fractions. The dependence of the solubility of a gas on its fugacity ensures that higher fractions of volatile elements are soluble in magma under real gas behaviour, thereby decreasing the availability of not only H, but also He and N in the envelope (Fig.~\ref{fig:full_model_ideal_real}(c)). The solubility of C is negligibly affected due to the relatively low solubility of methane, the dominant C-bearing gas species. The differential partitioning of H, He, C and N in magma results in non-solar abundances in the envelope. For instance, the near complete partitioning of N in magma (Fig.~\ref{fig:full_model_ideal_real}(c)) results in extremely low abundances of N$_2$ and NH$_3$ in the envelope (Fig.~\ref{fig:full_model_ideal_real}(b)). The high solubility of N occurs due to the low resulting oxygen fugacity of the system (between IW--6 and IW--4.3 for the ideal case or IW--6.2 and IW--5.7 for the real case, where IW is the iron-wüstite buffer), at which nitrogen exists predominantly as nitride (N$^{3-}$) in the silicate liquid \citep{LMH03,2022GeCoA.336..291D}. This is evident in Fig.~\ref{fig:full_model_ideal_real}(c) for the ideal case, in which higher oxygen fugacities at higher H mass fractions, due to higher H$_2$O/H$_2$ ratios (Eq. \ref{eq:H2H2O}), result in a decrease in N solubility.

\subsection{Effects of Mantle Melt Fraction and Accreted Gas Metallicity} \label{sec:resultsEnvelope2}

We calculate the MEB envelope composition of TOI-421b in the H--He--C--N--O--Si chemical system for accreted gas metallicities of 1$\times$ solar and 100$\times$ solar and mantle melt fractions of 100\% and 1\% (Sect.~\ref{sec:methodsHHeCNOSi}). Our results show both mantle melt fraction and accreted gas metallicity play key roles in controlling the envelope composition (Fig.~\ref{fig:full_model_envelope}). Although the real gas fugacity coefficients do not show extreme differences between the considered cases (Fig.~\ref{fig:full_model_fugcoeff}), the decreasing mantle melt fractions limit the ability of a given gas species to be accommodated in the liquid (Fig.~\ref{fig:full_model_solubility}). The differential partitioning of H, He, C and N (whose budgets are derived from the accreted gas) and the vaporisation of Si and O (whose budgets are largely based on the molten mantle) leads to final envelope budgets that differ from the total elemental budgets defined by the imposed metallicity.

For 0.1--1~wt\%~H and 100\% melt, MEB total pressures range between 13--77~kbar (1.3--7.7~GPa) at 1$\times$ solar and  between 11--69~kbar (1.1--6.9~GPa) at 100$\times$ solar (Fig.~\ref{fig:full_model_envelope}(a,b)). When the mantle melt fraction is set to 1\%, MEB total pressures range between 14--104~kbar (1.4--10.4~GPa) at 1$\times$ solar and between 17--129~kbar (1.7--12.9~GPa) at 100$\times$ solar (Fig.~\ref{fig:full_model_envelope}(c,d)). At 100\% melt and 1 $\times$ solar metallicity, magma vaporisation causes the partial pressures of SiO and SiH$_4$ to exceed that of H$_2$. At 1\% melt, the mass of magma and hence the Si and O budgets decrease by a factor of about 100, thereby decreasing the partial pressures of SiO and SiH$_4$, lowering the solubility of volatiles, and enriching the envelope in H$_2$ and He.

Increasing the accreted gas metallicity from 1$\times$ to 100$\times$ solar increases the MEB partial pressures of C- and N-bearing gases by a factor 100, whereas those of other gases remain largely unaffected (Fig.~\ref{fig:full_model_envelope}(a,b) and (c,d)).  At solar metallicity, the relatively low volatile inventories of C and N result in the envelope being essentially devoid of C- and N-bearing gases independent of melt fraction; the most abundant of these, CH$_4$, does not exceed mixing ratios of 1\%. By contrast, at 100\% melt, the high metallicity (100 $\times$ solar) case shows that CH$_4$ is the most abundant gas species across the majority of the H mass fraction range explored, while SiO (low H mass fraction), SiH$_4$ and H$_2$ (high H mass fraction) are also major constituents. At 100\% melt, low partial pressures of N-bearing gases occur due to the high solubility of N in magma (Fig.~\ref{fig:full_model_solubility}(a,b)) at low oxygen fugacities (between IW--6.2 and IW--5.2). Due to the low partitioning of C and N in the 1\% melt and 100$\times$ solar case at a higher oxygen fugacity of IW--2 (Fig.~\ref{fig:full_model_solubility}(d)), CH$_4$ and to a smaller extent N$_2$ become key species in the H$_2$- and He-rich envelope (Fig.~\ref{fig:full_model_envelope}(d)). Nonetheless, the sum of partial pressures of SiO and SiH$_4$ is comparable to those of H$_2$O (in the 1$\times$ solar case) or N$_2$ (in the 100$\times$ solar case).

\begin{figure*}
    \includegraphics[width=\textwidth]{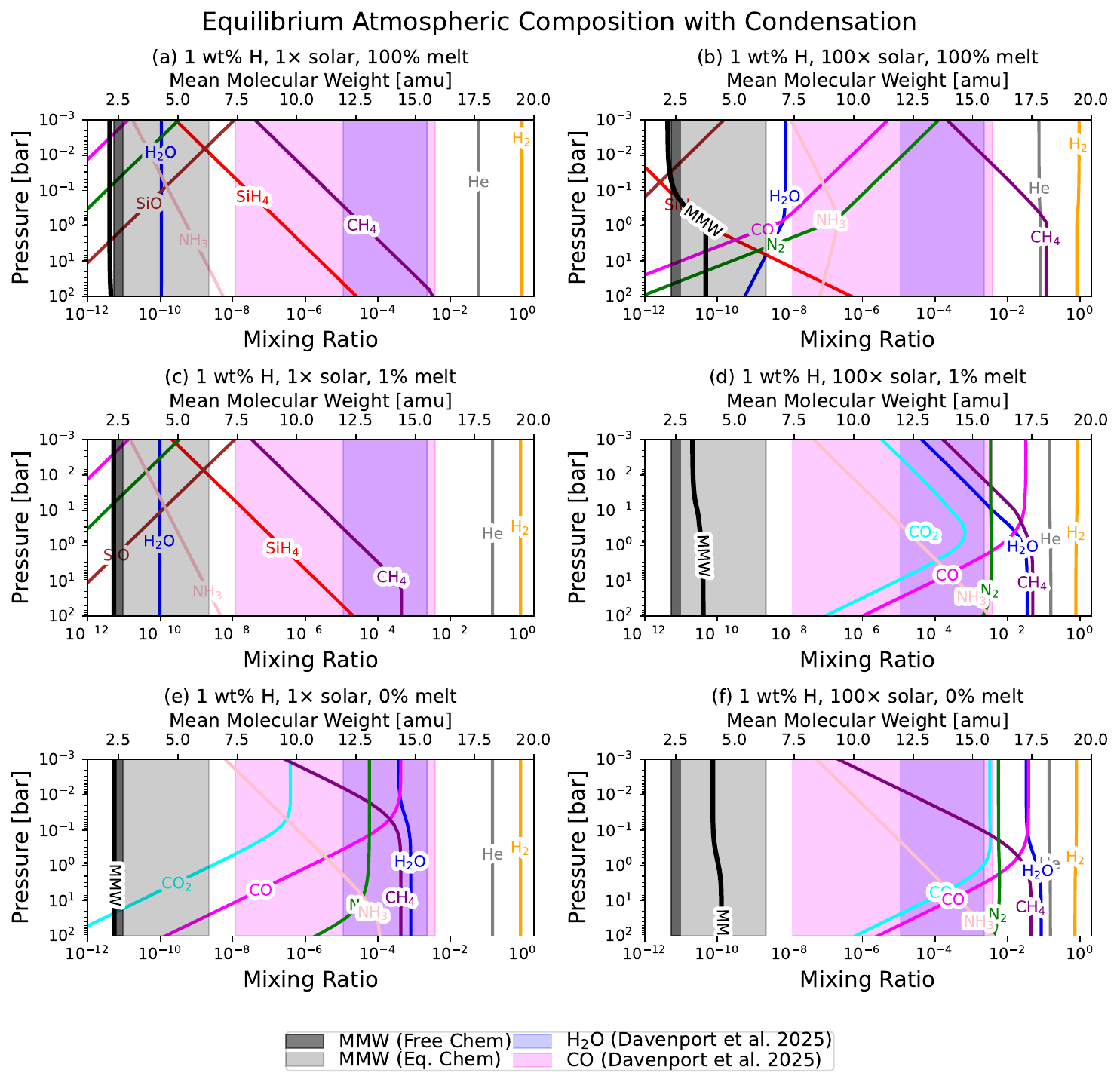}
    \vspace{-0.5cm}
    \caption{Atmospheric composition as a function of pressure (atmospheric depth) for TOI-421b under chemical equilibrium with envelope elemental abundances obtained from Fig.\ref{fig:full_model_envelope} with a melt fraction of 100\% (top panels), 1\% (middle panels), and another case with 0\% melt (bottom panels), not shown previously.  All calculations are performed at TOI-421b's equilibrium temperature, $T_{\rm eq}=920$~K, which is a reasonable approximation given that most parameterised $T-P$ profiles, within the pressure range of 1~mbar to 100~bar, are near-isothermal with a temperature close to $T_{\rm eq}$. Equilibrium condensation calculations of three polymorphs of SiO$_2$, Si, C, SiC and Si$_3$N$_4$ are included (Fig~\ref{fig:full_model_atmosphere_condensates}). The left and right panels are for cases with 1$\times$ and 100$\times$ solar metallicity, respectively. The shaded regions are observational constraints on the mixing ratios of H$_2$O and CO and the mean molecular weight (MMW) of TOI-421b from \citet{2025ApJ...984L..44D}.}   
    \label{fig:full_model_atmosphere}
\end{figure*}

\begin{figure*}
	\includegraphics[width=0.7\linewidth]{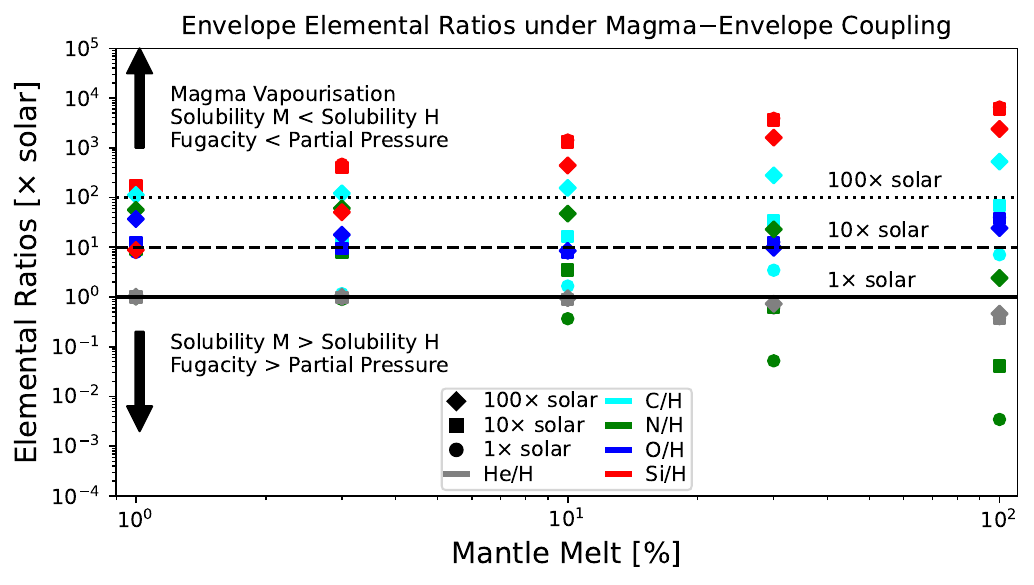}
    \caption{Envelope elemental ratios relative to H (in units of solar metallicity) of TOI-421b ($R_{\rm MEB} = 1.65 R_{\oplus}$, $R_P = 2.64 R_{\oplus}$, $M_P = 6.7 M_{\oplus}$, $T_{\rm MEB}=3000$~K, 1~wt\%~H) for the metallicity of accreted gas between 1$\times$, 10$\times$ and 100$\times$ solar and mantle melt fraction between 1--100\%. }
    \label{fig:summary}
\end{figure*}

\subsection{Equilibrium Atmospheric Composition with Condensation} \label{sec:resultsAtmosphere}

The magma--envelope coupling results in elemental abundances in the envelope that differ from their budgets in the accreted gas due to a differential solubility of considered gases (Sect.~\ref{sec:resultsEnvelope2}, Fig.~\ref{fig:full_model_solubility}). These envelope elemental abundances for the 100\% and 1\% melt cases are used as input to \texttt{Atmodeller} to evaluate the variation of atmospheric composition with atmospheric pressure (1~mbar--100~bar). To simulate the effect of atmospheric pressure, all elemental mass constraints are scaled down by the same factor, while keeping the elemental abundances relative to each other fixed at their MEB values. The reaction network presented in Sects. \ref{sec:resultsEnvelope1} and \ref{sec:resultsEnvelope2} (Table~\ref{tab:reactions}) is expanded to include seven condensates, three SiO$_2$ polymorphs, Si, C and SiC, Si$_3$N$_4$ (Appendix~\ref{app:condensation}). The network is solved at a fixed temperature $T_{\rm eq}=920$~K. The SiO$_2$ condensation ($\alpha$-quartz) reaction \citep{MZG02}, which is stoichiometrically the same as the magma--gas reaction (Eq. \ref{eq:SiO2}), replaces SiO$_{\rm 2(l)}$ with SiO$_{\rm 2(\alpha-Qz)}$. For cases without magma (mantle melt fraction = 0), the envelope elemental abundances are defined by the metallicity of the accreted gas. For all cases, we assume that the envelope contains 1~wt\% H relative to planet mass, an observationally inferred constraint from the mass and radius of TOI-421b \citep{2025ApJ...984L..44D}. The resulting atmosphere behaves ideally (a reasonable approximation at atmospheric pressures between 1~mbar and 100~bar), and is well-mixed (i.e., elemental abundances remain constant as a function of atmospheric pressure).

Equilibrium condensation of $\alpha$-quartz, a SiO$_2$ polymorph, occurs in the atmosphere when the mantle melt fraction is 1--100\% across all metallicities (Fig.~\ref{fig:full_model_atmosphere}(a--d)). The condensation of silicate species such as quartz is crucial for the formation of silicate clouds \citep{2024Natur.625...51D,2025Hoch}. For solar metallicity cases, SiC and Si also condense. In contrast, for 100$\times$ solar cases, C (graphite) condenses together with $\alpha$-quartz. Furthermore, SiC also condenses for the 100$\times$ solar and 100\% melt case. 
Even after the removal of a large fraction of elemental Si and C from the gas phase, for the solar metallicity cases, SiH$_4$, SiO and CH$_4$ remain as key species with mixing ratios below 10$^{-4}$ in the 1~mbar--1~bar range. For 100$\times$ solar cases, Si-bearing gases exhibit 2--4 orders of magnitude lower mixing ratios at 100\% melt and are absent at 1\% melt, and CH$_4$ exhibits 4 orders of magnitude higher mixing ratios. For the case with 100$\times$ solar metallicity and 1\% melt, the H$_2$O mixing ratio is the highest among all considered cases. In the absence of magma, mixing ratios of Si-bearing gases are significantly lower than those of C-bearing gases (Fig.~\ref{fig:full_model_atmosphere}(e,f)), as the nebular gas is the only source of Si in this case. 
At higher atmospheric pressures, CH$_4$ and SiH$_4$ exhibit higher mixing ratios than CO and SiO, respectively, because increasing pressure favours the forward reactions (Eqs.~\ref{eq:silane_fm} and \ref{eq:methane_fm}) with fewer gas molecules, consistent with Le Chatelier’s principle.

\section{Observability of Magma--Envelope Coupling} \label{sec:implications}

Building on the compositional models presented in Section~\ref{sec:results}, we now interpret the results in the context of observations and identify potential atmospheric diagnostics of magma--envelope coupling. As shown in Fig.~\ref{fig:summary}, the final envelope elemental abundances relative to H diverge strongly from those expected purely from accreted nebular gas, particularly at high mantle melt fractions. This divergence arises from four key factors. 
First, higher mantle melt fractions result in higher Si budgets in the system and, consequently, increase the partial pressures of Si-bearing gases in the envelope.
Second, mantle melt fractions approaching 1\% and accreted gas metallicities approaching 100$\times$ solar produce more oxidising conditions ($\log_{10}f_{\rm O_2}$~$\sim$~IW--2) than for other cases ($\log_{10}f_{\rm O_2}$~$\sim$~IW--6). Our models self-consistently calculate the oxygen fugacity with the total O budget as an elemental mass constraint given the available equilibrium constants of reactions (Table \ref{tab:reactions}). More accurate estimates of oxygen fugacity will require constraints on H--O--Si--Fe equilibria at several GPa, for which experimental constraints are presently lacking.
%following global chemical equilibrium models \citep{2022PSJ.....3..127S,2025ApJ...988L..55W}. However, it remains unclear whether using oxygen fugacity as a constraint is more suited to the magma--envelope boundary of sub-Neptunes \citep{2025ApJ...987..174I,2025ApJ...994...28H,2025arXiv250700499B}. }
Third, the behaviour of real gas species can result in positive or negative deviations from ideality, leading to lower or higher partial pressures of these gases, respectively. Specifically, SiO exhibits a negative deviation from ideality, in contrast with all other considered gases. 
Fourth, the solubility of gases in the melt depletes volatile elements (H, He, C, N) in the envelope to varying extents. Elemental abundances relative to H (in units of metallicity) can therefore increase or decrease compared to the accreted gas metallicity. Reducing conditions at mantle melt fraction of 100\% ($\log_{10}f_{\rm O_2}$ $\sim$ IW--6) preferentially enhance the solubility of N over H and especially over C. In contrast, relatively oxidising conditions at mantle melt fraction of 1\% and 1~wt\%~H ($\log_{10}f_{\rm O_2}$ $\sim$ IW--2) result in a relatively lower solubility of N (Fig.~\ref{fig:full_model_solubility}(c)). Consequently, envelopes in contact with molten mantles are predicted to be strongly depleted in N and weakly depleted in C, with enhanced abundances of Si and O. 

A notable exception is the formation of silane SiH$_4$ via the reaction
\begin{equation}
    \mathrm{SiO + 3H_2 \rightleftharpoons SiH_4 + H_2O},
    \label{eq:silane_fm}
\end{equation}
requiring both the magma-derived Si and the envelope-derived H. Because the generation of SiH$_4$ in the envelope is favoured at higher pressures, the dissolution of H$_2$ in the magma ocean is hindered. The analogous reaction describes the formation of methane CH$_4$, where Si is replaced by C in the reaction
\begin{equation}
    \mathrm{CO + 3H_2 \rightleftharpoons CH_4 + H_2O}.
    \label{eq:methane_fm}
\end{equation}
Both reactions are pressure-dependent and exhibit stark non-ideal behaviour at the magma-envelope boundary of sub-Neptunes (Sect.~\ref{sec:methodsRealGas}). Because H$_2$ also becomes more soluble due to an increase (albeit more mild) in its fugacity coefficient, the resulting increase in both Si/H and C/H ratios in the envelope nevertheless stabilises methane and silane relative to the ideal case. The competition between SiH$_4$ and CH$_4$ in the envelope is therefore controlled by the mantle melt fraction, which largely governs the abundance of Si in the envelope, and the metallicity of the accreted gas, which governs the abundance of C in the envelope.

\begin{figure}
	\includegraphics[width=\linewidth]{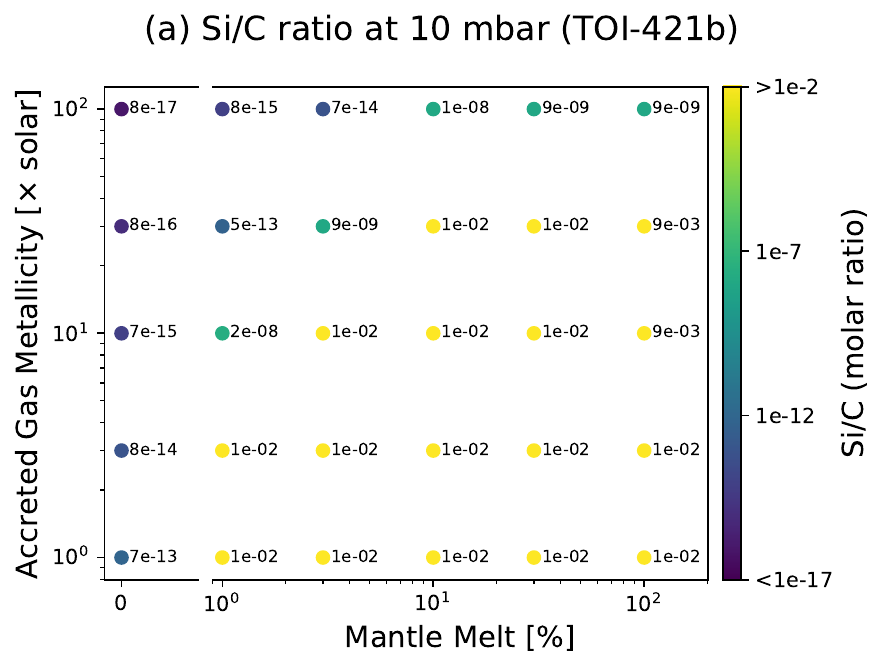}
    \includegraphics[width=\linewidth]{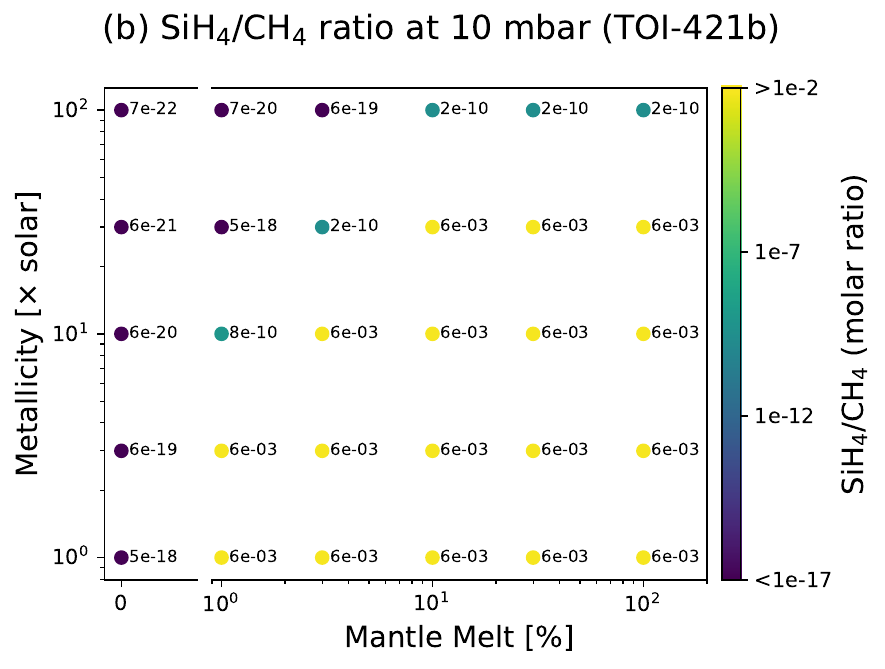}
    \caption{Gas-phase Si/C (a) and SiH$_4$/CH$_4$ ratios (b) at the pressure of 10~mbar in the atmosphere of TOI-421b ($R_{\rm MEB} = 1.65 R_{\oplus}$, $R_P = 2.64 R_{\oplus}$, $M_P = 6.7 M_{\oplus}$, $T_{\rm MEB}=3000$~K, $T_{\rm eq}=920$~K, 1~wt\% H) for metallicities of accreted gas between 1--100$\times$ solar and mantle melt fractions of 0\% and 1--100\%. }
    \label{fig:full_model_SiCratios}
\end{figure}

\begin{figure}
	\includegraphics[width=\linewidth]{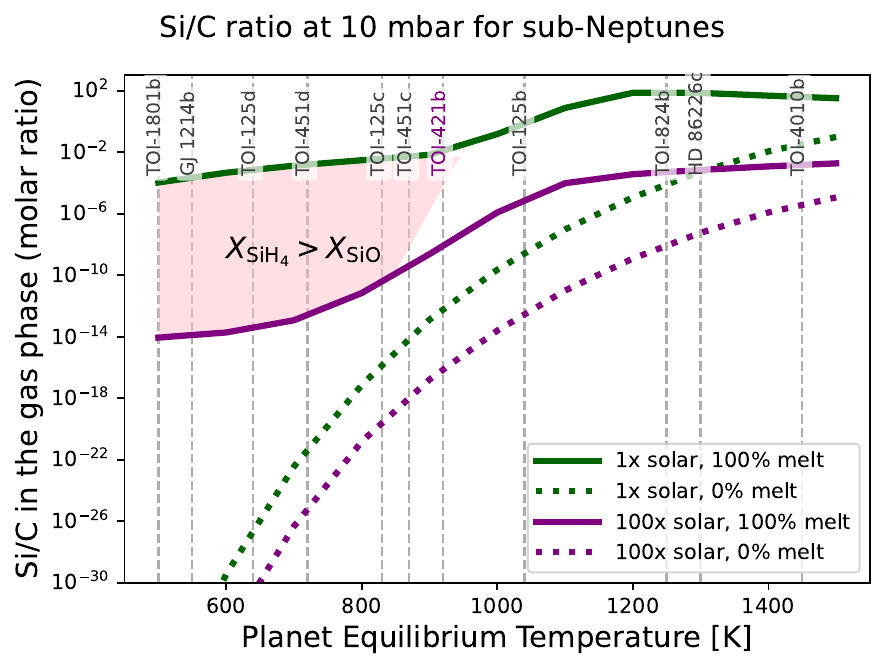}
    \caption{Si/C ratios in the gas phase at 10~mbar for a sub-Neptune ($R_{\rm MEB} = 1.65 R_{\oplus}$, $R_P = 2.64 R_{\oplus}$, $M_P = 6.7 M_{\oplus}$, $T_{\rm MEB}=3000$~K, 1~wt\% H) as a function of planet equilibrium temperature $T_{\rm eq}$ (= atmospheric temperature $T_{\rm atm}$), representing several sub-Neptunes that are \textit{JWST} targets. Volume mixing ratios of SiH$_4$ $X_{\rm SiH_4}$ exceed that of SiO $X_{\rm SiO}$ for $T_{\rm eq} <900$~K, which is highlighted by the shaded region. TOI-421b is at the transition with $X_{\rm SiH_4} \sim X_{\rm SiO}$.}
    \label{fig:full_model_SiCtrends}
\end{figure}

Our models at 10~mbar pressures show that the relative mixing ratio of SiH$_4$ and CH$_4$ and the Si/C ratio in general (Fig.~\ref{fig:full_model_SiCratios}), are strong diagnostics of magma--envelope coupling. We find that SiH$_4$/CH$_4$ $\sim$ Si/C $\sim$ 0.01 serves as an indicator for observationally identifying magma–envelope coupling in TOI-421b. This threshold is consistently met at high mantle melt fractions, especially at low to moderate metallicities. Si- and C-bearing condensates limit the gas-phase ratios to an upper bound (Sect.~\ref{sec:resultsAtmosphere}). For equilibrium temperatures below $\sim$900~K and fully molten magma oceans, SiH$_4$ is the predominant Si-bearing species in the atmosphere, regardless of the accreted gas metallicity (Fig.~\ref{fig:full_model_SiCtrends}). On the other hand, for equilibrium temperatures exceeding 1000~K, SiO is the predominant Si-bearing species and the differences between 100\% melt and 0\% melt cases are not readily discernible. TOI-421b lies in the transition zone, making both SiO and SiH$_4$ potentially detectable. Therefore, the detection of elevated SiH$_4$ abundances in sub-Neptunes with atmospheric temperatures below $\sim$1000~K could provide observational evidence for ongoing or past magma–envelope coupling.

Comparing our predictions with the observed transmission spectrum of TOI-421b, the retrieved mixing ratios of H$_2$O and CO lie between 10--1000 ppm and 0.01--1000 ppm, respectively, with a mean molecular weight (MMW) of 2.3--2.7~amu for free chemistry retrievals and up to 6.3~amu for equilibrium chemistry retrievals \citep{2025ApJ...984L..44D}. These constraints are consistent with our model for a 1$\times$ solar atmosphere without magma (0\% melt; Fig.~\ref{fig:full_model_atmosphere}(e)), but are also broadly matched by our 1\% melt, 100$\times$ solar case (Fig.~\ref{fig:full_model_atmosphere}(d), exhibiting more oxidising conditions than other magma-bearing cases), except for methane, which is predicted to be more abundant than observed. This discrepancy may be due to disequilibrium processes, such as photochemical depletion of CH$_4$, which are not included in our equilibrium models \citep{2011ApJ...737...15M}.

Current retrievals for TOI-421b and sub-Neptunes with equilibrium temperatures lower than 1000~K do not include Si-bearing species, and thus cannot constrain the Si/C ratio observationally. We recommend that future retrievals incorporate species such as SiO and SiH$_4$, which are potentially detectable in the infrared \citep{2017MNRAS.471.5025O,2023A&A...671A.138Z,2025ApJ...987..174I}. Detections of both Si- and C-bearing gases would allow for the inference of Si/C ratios and hence the mantle melt fraction and the metallicity of the accreted gas.

Although sub-Neptune metallicities inferred from population-level mass–metallicity relations tend to be high, observational constraints span a wide range from solar to 100$\times$ solar \citep{2023ApJ...956L..13M,2024arXiv240303325B,2025ApJ...984L..44D}. For example, V1298~Tau~b, a hot sub-Neptune likely in an earlier evolutionary phase than TOI-421b, exhibits a super-solar metallicity, but still below the population trend \citep{2025arXiv250708837B}. Free-chemistry retrievals treat metallicity as a derived parameter, which complicates comparison across different retrieval approaches. Incorporating Si-bearing species and adopting consistent chemical frameworks across retrieval methods \citep{2025A&A...701A.133K} will be important for placing robust constraints on magma–envelope interactions.

Future missions such as ARIEL, which will likely observe hundreds of sub-Neptunes with radii between 1.8--4~$R_\oplus$ \citep{2022AJ....164...15E}, offer promising opportunities to test these hypotheses. Joint modelling of magma–envelope coupling, equilibrium/disequilibrium chemistry, and atmospheric mass loss will be critical for interpreting atmospheric compositions and for constraining the thermochemical state of sub-Neptunes. Moreover, coupling the results of static compositional models with those of evolutionary mass loss of elements \citep{2025ApJ...991..121L} will allow comprehensive interpretations of volatile retention and atmospheric evolution in sub-Neptune populations.

\section{Conclusions}

We demonstrate that magma--envelope coupling, gas solubility and real gas equations of state, including for SiH$_4$ and SiO, substantially alter the chemical composition of hot sub-Neptune atmospheres from their primordial states. High mantle melt fractions drive silicate magma vaporisation, enriching the envelope in Si and O while depleting it in H, N, He and C via magma dissolution. In contrast, low melt fractions preserve more H$_2$- and He-rich compositions, even at high (100$\times$ solar) accreted gas metallicities, with C- and Si-bearing gases being present in subordinate abundances.  When mantle melt fractions are high, sufficient H dissolves in the magma ocean to increase Si/H and C/H ratios, leading to a competition between SiH$_4$ at low metallicity and CH$_4$ at high metallicity. This interplay makes high silane/methane ratios a promising diagnostic of magma--envelope coupling sub-Neptunes. Even after considering the condensation of `clouds' of $\alpha$-quartz, graphite, metallic silicon and silicon carbide, the resulting silane/methane and Si/C ratios in the upper atmosphere at pressures relevant to transmission spectroscopy measurements, reflect the nature of the underlying magma ocean and the metallicity of the accreted nebular gas. This highlights their role in controlling the observable atmospheric composition for sub-Neptunes with equilibrium temperatures lower than 1000~K. In contrast, the presence of H$_2$- and He-rich, SiH$_4$-deficient and CH$_4$-bearing atmospheres may indicate a limited role or absence of magma oceans on sub-Neptunes.

%%%%%%%%%%%%%%%%%%%%%%%%%%%%%%%%%%%%%%%%%%%%%%%%%%
\section*{Acknowledgements}

KH acknowledges the FED-tWIN research program STELLA (Prf-2021-022) funded by the Belgian Science Policy Office (BELSPO), the Research Foundation Flanders (FWO) research grant G014425N, and COST Action CA22133 PLANETS. Part of this work is supported by the Belgo-Indian Network for Astronomy and astrophysics (BINA), approved by the International Division, Department of Science and Technology (DST, Govt. of India; DST/INT/BELG/P-09/2017) and BELSPO (Govt. of Belgium; BL/33/IN12). PAS, DJB and FLS were supported by the Swiss State Secretariat for Education, Research and Innovation (SERI) under contract No. MB22.00033, a SERI-funded ERC Starting grant ``2ATMO''. PAS also thanks the Swiss National Science Foundation (SNSF) through an Eccellenza Professorship (\#203668). KH thanks Leen Decin, Linus Heinke, Thomas Konings, Attilio Rivoldini, Tim Van Hoolst, Denis Defr\'{e}re and Olivier Namur for insightful discussions. We thank an anonymous reviewer for their valuable comments that led to crucial improvements to the manuscript.

%%%%%%%%%%%%%%%%%%%%%%%%%%%%%%%%%%%%%%%%%%%%%%%%%%
\section*{Data Availability}
All data were produced using the open-source software package \texttt{Atmodeller} v1.0.0 (\url{https://github.com/ExPlanetology/atmodeller}). The Python notebooks to generate data and the data used in this study are available under the GNU General Public License (GPL) 3.0 and can be downloaded \citep{2025HakimData}.
 
%The inclusion of a Data Availability Statement is a requirement for articles published in MNRAS. Data Availability Statements provide a standardised format for readers to understand the availability of data underlying the research results described in the article. The statement may refer to original data generated in the course of the study or to third-party data analysed in the article. The statement should describe and provide means of access, where possible, by linking to the data or providing the required accession numbers for the relevant databases or DOIs.

%%%%%%%%%%%%%%%%%%%% REFERENCES %%%%%%%%%%%%%%%%%%

% The best way to enter references is to use BibTeX:

\bibliographystyle{mnras}
\bibliography{hakim} 

%%%%%%%%%%%%%%%%%%%%%%%%%%%%%%%%%%%%%%%%%%%%%%%%%%

%%%%%%%%%%%%%%%%% APPENDICES %%%%%%%%%%%%%%%%%%%%%

\appendix

\section{Full Range of MEB Partial Pressures} \label{app:fullrange}

\begin{figure*}
	\includegraphics[width=\textwidth]{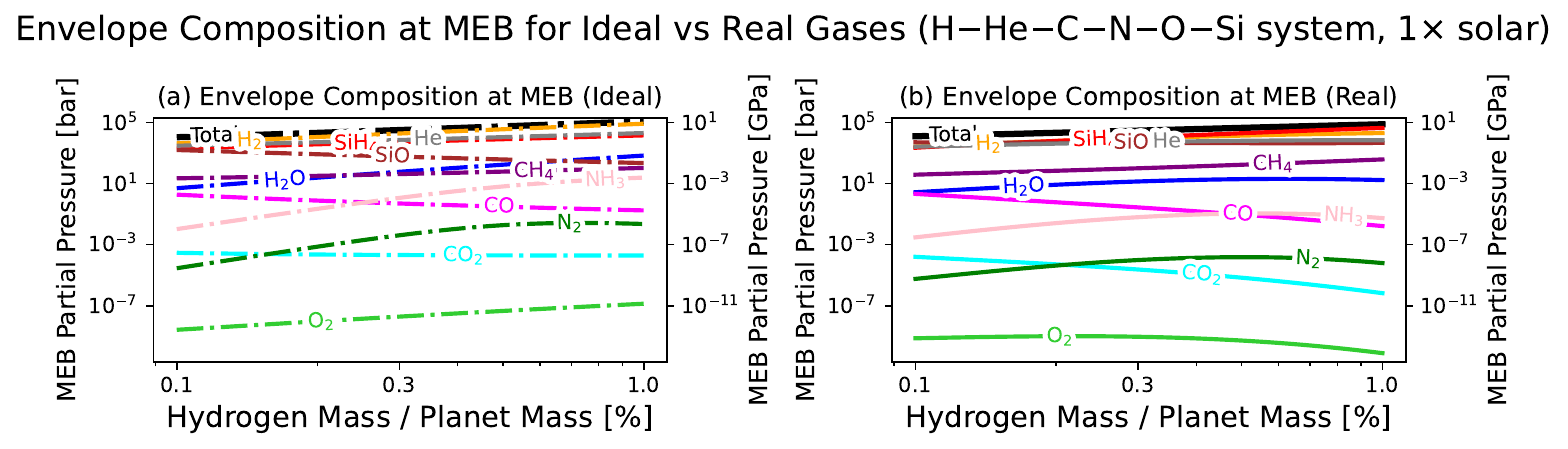}
    \caption{The same as Fig.~\ref{fig:full_model_ideal_real}(a,b) but partial pressures of all gases shown with vertical axes spanning 16 orders of magnitude. }
    \label{fig:full_model_ideal_real_fullrange}
\end{figure*}

\begin{figure*}
	\includegraphics[width=\textwidth]{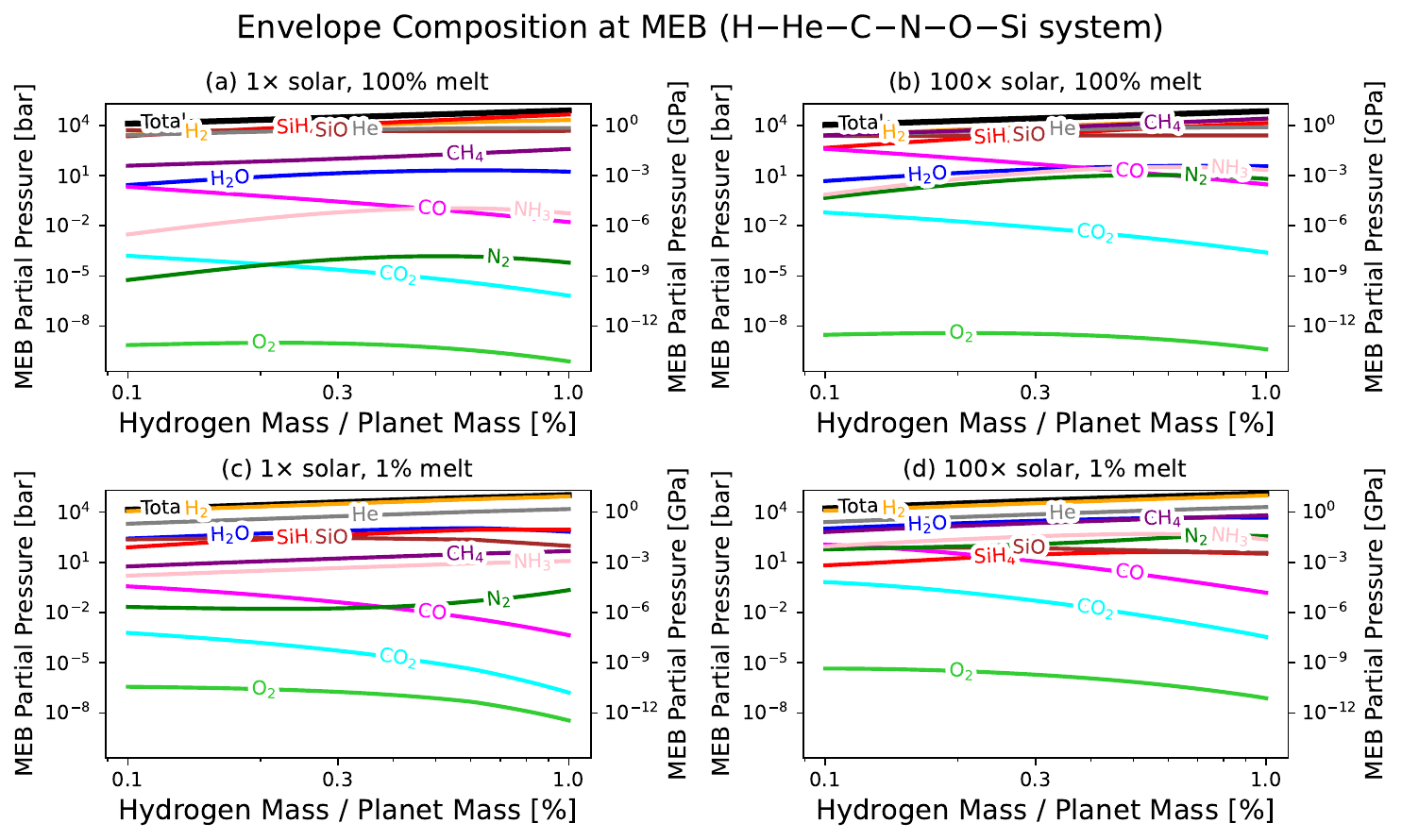}
    \caption{The same as Fig.~\ref{fig:full_model_envelope} but partial pressures of all gases shown with vertical axes spanning 16 orders of magnitude. }
    \label{fig:full_model_envelope_fullrange}
\end{figure*}

Sections \ref{sec:resultsEnvelope1} and \ref{sec:resultsEnvelope2} only show gases with MEB partial pressures exceeding 1~bar for clarity. For completeness, Fig.~\ref {fig:full_model_ideal_real_fullrange} and Fig.~\ref {fig:full_model_envelope_fullrange} present the full range of calculated MEB partial pressures of all considered gases in the H--He--C--N--O--Si chemical system for TOI-421b. Fig.~\ref {fig:full_model_ideal_real_fullrange}(a,b) shows the same set of models as Fig.~\ref{fig:full_model_ideal_real}(a,b) with the full pressure range. Fig.~\ref {fig:full_model_envelope_fullrange} shows the same set of models as Fig.~\ref{fig:full_model_envelope} but with the full range of MEB partial pressures. 

\section{Element Partitioning and Fugacity Coefficients} \label{app:solubility_fugacity}

\begin{figure*}
	\includegraphics[width=0.9\textwidth]{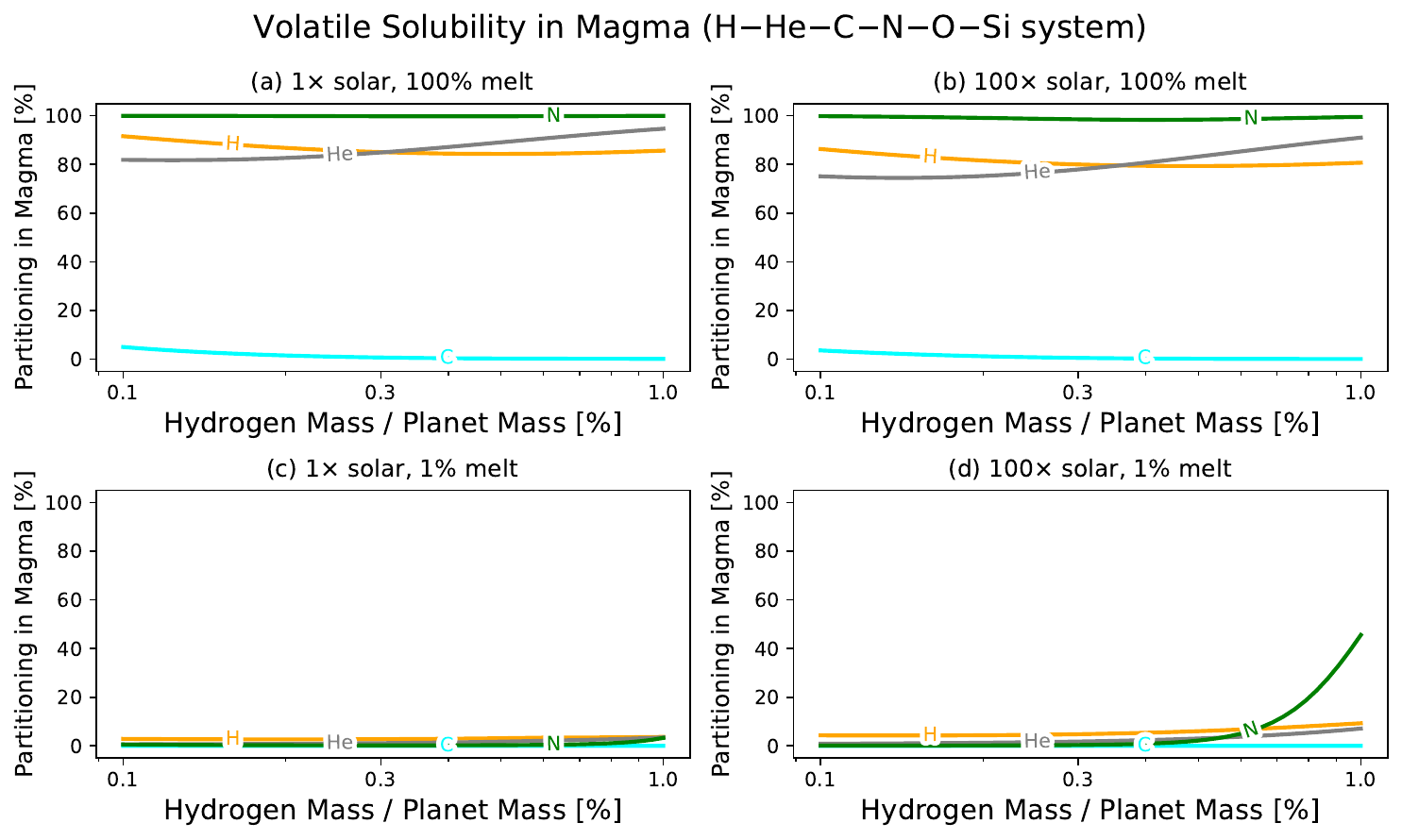}
    \caption{Volatile element partitioning in magma in the H--He--C--N--O--Si chemical system of TOI-421b corresponding to the envelope compositions in Fig.~\ref{fig:full_model_envelope}. }
    \label{fig:full_model_solubility}
\end{figure*}

\begin{figure*}
	\includegraphics[width=0.9\textwidth]{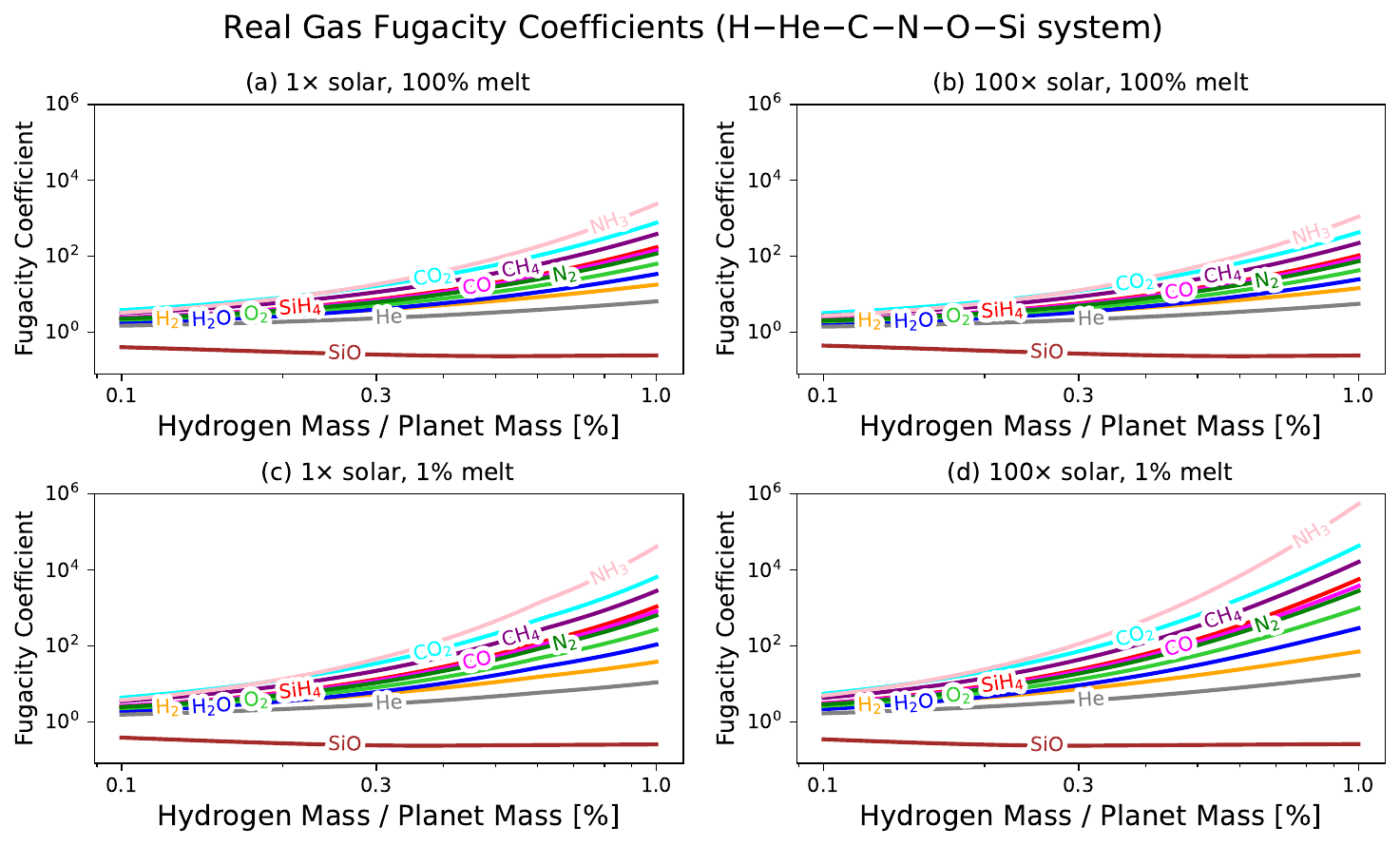}
    \caption{Real gas fugacity coefficients in the H--He--C--N--O--Si chemical system of TOI-421b corresponding to the envelope compositions in Fig.~\ref{fig:full_model_envelope}. }
    \label{fig:full_model_fugcoeff}
\end{figure*}

Fig.~\ref{fig:full_model_solubility} and \ref{fig:full_model_fugcoeff} respectively show the volatile element partitioning in magma and the real gas fugacity coefficients for the magma--envelope compositions of TOI-421b for the four end-member cases with mantle melt fractions of 1\% and 100\% and the accreted gas metallicities of 1$\times$ and 100$\times$ solar. 

\section{Condensation of Si- and C-bearing Species in the Atmosphere} \label{app:condensation}

\begin{figure*}
	\includegraphics[width=0.9\textwidth]{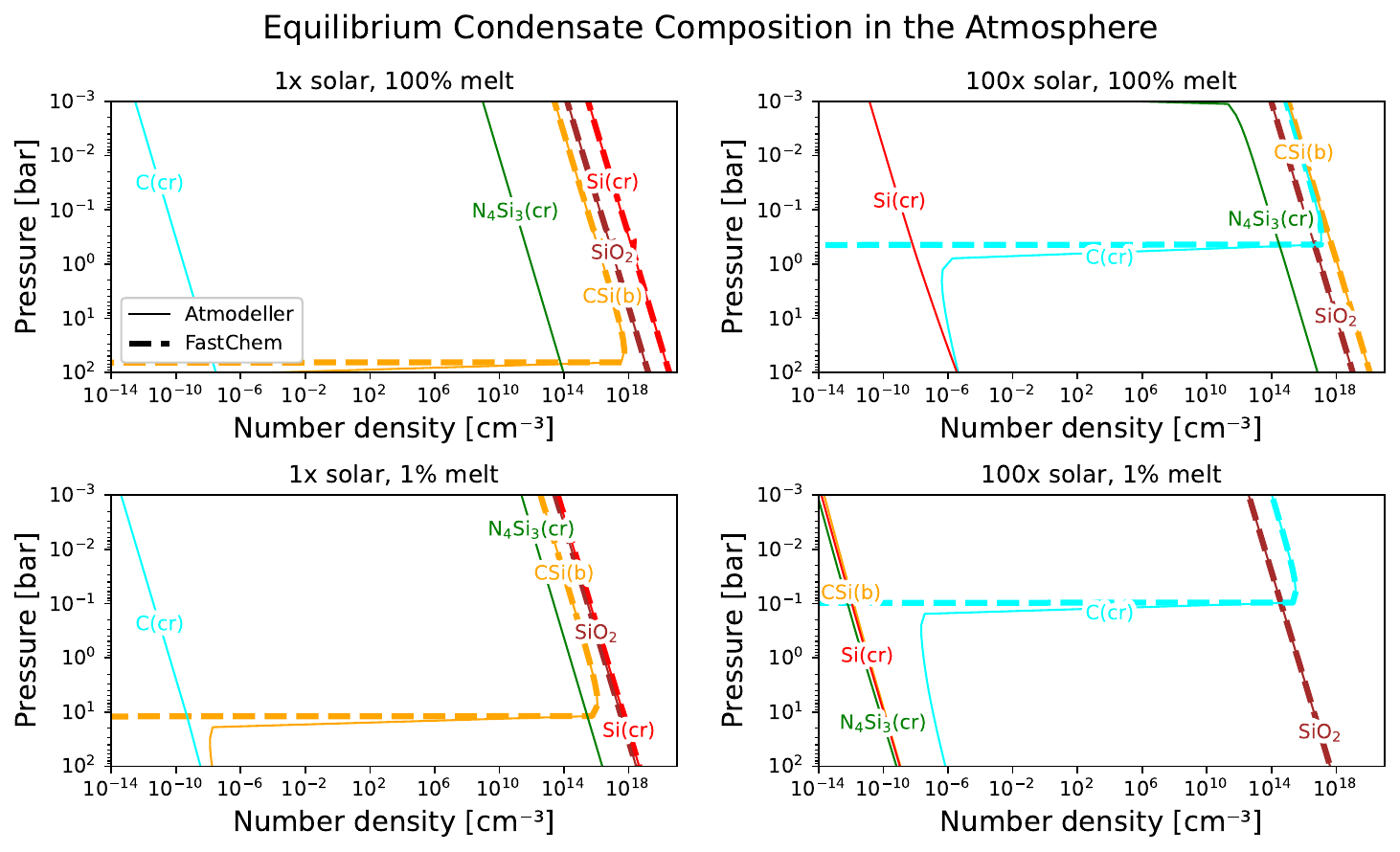}
    \caption{Condensates for the four cases with non-zero mantle melt fractions from Fig.~\ref{fig:full_model_atmosphere}. The comparison between \texttt{Atmodeller} \citep{2025arXiv250700499B} and \texttt{FastChemCOND} \citep{2024MNRAS.527.7263K} is shown. The SiO$_2$ number density is controlled by $\alpha$-quartz. Si$_3$N$_4$ is not considered in the \texttt{FastChemCOND} calculation. }
    \label{fig:full_model_atmosphere_condensates}
\end{figure*}

Seven additional chemical reactions describing the condensation of seven Si- and C-bearing species are considered in atmosphere models compared to models at the magma--envelope boundary. The magma--gas reaction (Reaction 6, Table~\ref{tab:reactions}) is not considered in atmosphere models. The reactions involving the condensates, C, Si, SiC, Si$_3$N$_4$ and three SiO$_2$ polymorphs ($\alpha$-Quartz, $\beta$-Quartz and $\beta$-Cristobalite), are listed in Table~\ref{tab:reactionsCondensation}. As described in Sect.~\ref{sec:methods}, \texttt{Atmodeller} applies the extended law of mass action to evaluate the stability of all condensates and partitioning of elements between gaseous and condensed phases \citep{2025arXiv250700499B}. If a condensate is stable, its activity is unity. The condensation of Si- and C-bearing species for the models from Fig.~\ref{fig:full_model_atmosphere_condensates} exhibits negligible differences in number densities exceeding 1~cm$^3$ between \texttt{Atmodeller} and \texttt{FastChemCOND} \citep{2024MNRAS.527.7263K}, demonstrating the validity of our results. The differences below 1~cm$^3$ between \texttt{Atmodeller} and \texttt{FastChemCOND} are likely due to our consideration of an additional condensate, Si$_3$N$_{4}$ and the tolerance values for numerical convergence. SiO$_{\rm 2(\alpha-Qz})$ is the universal condensate in all models. The degree of condensation of elemental Si and C is also shown.

\begin{table*}
\centering
\caption{Condensation reactions in the H--He--C--N--O--Si chemical system} \label{tab:reactionsCondensation}
\begin{tabular}{rclcl}
\hline
\hline
\multicolumn3c{Chemical Reactions} & Reaction Constant & Reference  \\
\hline
\multicolumn{3}{l}{Condensation Reactions} \\
\(\rm{SiO_{2(\alpha-Qz)}}\) & \(\rightleftharpoons\) & \(\rm{SiO_{(g)}  + 0.5\,O_{2(g)}}\) & $K_{14} = \frac{f_{\rm{SiO_{(g)}}} \sqrt{f_{\rm{O_{2(g)}}}} }{a_{\rm{SiO_{2(\alpha-Qz)}}}}$ & \citet{MZG02} \\
\(\rm{SiO_{2(\alpha-Qz)}}\) & \(\rightleftharpoons\) & \(\rm{SiO_{2(\beta-Qz)}}\) & $K_{15} = \frac{a_{\rm{SiO_{2(\beta-Qz)}}}}{a_{\rm{SiO_{2(\alpha-Qz)}}}}$ & \citet{MZG02} \\
\(\rm{SiO_{2(\beta-Crt)}}\) & \(\rightleftharpoons\) & \(\rm{SiO_{2(\beta-Qz)}}\) & $K_{16} = \frac{a_{\rm{SiO_{2(\beta-Qz)}}}}{a_{\rm{SiO_{2(\beta-Crt)}}}}$ & \citet{MZG02} \\
\(\rm{C_{(cr)} + 2\,H_{2(g)}}\) & \(\rightleftharpoons\) & \(\rm{CH_{4(g)}}\) & $K_{17} = \frac{f_{\rm{CH_{4(g)}}}}{a_{\rm{C_{(cr)}}} f^2_{\rm{H_{2(g)}}} }$ & \citet{MZG02} \\
\(\rm{Si_{(cr)} + 2\,H_{2(g)}}\) & \(\rightleftharpoons\) & \(\rm{SiH_{4(g)}}\) & $K_{18} = \frac{f_{\rm{SiH_{4(g)}}}}{a_{\rm{Si_{(cr)}}} f^2_{\rm{H_{2(g)}}} }$ & \citet{MZG02} \\
\(\rm{Si_{(cr)} + C_{(cr)}}\) & \(\rightleftharpoons\) & \(\rm{SiC_{(cr)}}\) & $K_{19} = \frac{a_{\rm{SiC_{(cr)}}}}{a_{\rm{Si_{(cr)}}} a_{\rm{C_{(cr)}}} }$ & \citet{MZG02} \\
\(\rm{Si_3N_{4(cr)}}\) & \(\rightleftharpoons\) & \(\rm{3 Si_{(cr)}  + 2\,N_{2(g)}}\) & $K_{20} = \frac{a_{\rm{Si_{(cr)}}} f^2_{\rm{N_{2(g)}}} }{a_{\rm{Si_3N_{4(cr)}}}}$ & \citet{MZG02} \\

\hline
\end{tabular}
\end{table*}

% Don't change these lines
\bsp	% typesetting comment
\label{lastpage}
\end{document}